\documentclass[12pt]{iopart}
\usepackage[utf8]{inputenc}
\usepackage[T1]{fontenc}
\usepackage[dvipsnames]{xcolor}
\usepackage{amssymb,bm,graphicx}

\newcommand{\dfrac}[2]{\displaystyle\frac{\displaystyle #1}{\displaystyle #2}}

\begin{document}

\title[Approximate self-energy for Fermi systems with large s-wave scattering length]{Approximate self-energy for Fermi systems with large s-wave scattering length: a step towards density functional theory.}
\author{Antoine Boulet$^{1}$, and Denis Lacroix$^{1}$}
\address{$^1$Institut de Physique Nucl{\'e}aire, IN2P3-CNRS, Universit{\'e} Paris-Sud,Universit{\'e} Paris-Saclay, F-91406 Orsay Cedex, France}

\begin{abstract}
In the present work, we start from a minimal Hamiltonian for Fermi systems where the s-wave scattering is the only low energy constant at play. Many-Body Perturbative approach that is usually valid at rather low density is first discussed.
We then use the resummation technique with the ladder approximation to obtain compact expressions for both the energy and/or
the on-shell self-energy in infinite spin-degenerated systems. Diagrammatic resummation technique has the advantage in general to be predictive
in a region of density larger compared  to many-body perturbation theory. It also leads to non-diverging limit as $|a_s| \rightarrow + \infty$.
Still, the obtained expressions are rather complex functional of the Fermi momentum
$k_F$.  We introduce the
full phase-space average or the partial phase-space methods respectively applied to the energy or to the self-energy to simplify their  dependences
in terms of $(a_s k_F)$ while keeping the correct limit at low density and the non-diverging property at large $|a_s k_F|$.
Quasi-particle properties of Fermi system in various regime of density and scattering
length are then illustrated.  Our conclusion is that such simplified expressions where the direct link is made with the low energy constant
without fine-tuning can provide a clear guidance to obtain density functional theory beyond the perturbative regime. However, quasi-particle
properties close or near unitary cannot be reproduced unless this limit is explicitly used as a constraint.   We finally discuss how such approximate
treatment of quasi-particle can guide the development of density functional theory for strongly interacting
Fermi systems.
\end{abstract}

\maketitle
\section{Introduction}
Strongly interacting many-body Fermi systems can sometimes be described by rather simple density functional theories (DFT). This is the case
of nuclear systems where simple functionals, like those based on the Skyrme type contact interactions \cite{Sky59,Vau72,Ben03,Sto07}, are nowadays widely used.
With very few parameters,
the functional can describe static, thermodynamical and dynamical properties very accurately in a unified framework. This is even more surprising in
view of the complexity of the strong multi-body interaction between nucleons. The question (a) {\it ``How such simplicity can emerge in strongly interacting Fermi systems?''} is still largely
open in the nuclear physics community (see discussion in \cite{Sch12a}).

This situation is not unique in nature. Simple DFTs apply also to the case of Fermi gas at unitarity. This gases are characterized by infinite s-wave scattering length $a_s$ in the dilute regime. In this
case, the energy becomes directly proportional to the free Fermi-Gas energy. This situation can be seen as one of the simplest DFT one could ever imagine.  Still, while in a DFT framework unitary gases can be described in a rather simplistic manner (see for instance \cite{Pap05,Bul07,Lac16a}), their treatment starting from a particle-particle interactions,
requires rather advanced many-body techniques like Monte-Carlo (MC) methods \cite{Cha04a,Bul08a,For11a,Gez08a,Car12a,For12a,End13}, Self-Consistent Green Function (SCGF) \cite{Hau07a,Hau09a}, Brueckner Hartree-Fock (BHF) \cite{Dog14}, Blod Diagrammatic Monte Carlo \cite{Hou12a,Hou13a} eventually associated to resummation technique based on conformal-Borel transformation \cite{Ros18}. These approaches generally rely on rather involved numerical methods and usually prevent from connecting analytically the energy with the low-energy constants (LEC) associated to the underlying interaction. To our opinion, to reply to the question (a) it would be desirable to also give some hints on the other question (b) {\it ``Can we qualitatively or quantitatively connect the parameters of the DFTs used in strongly interacting Fermions with the low-energy constants of the interaction?''}
Assuming that we can directly connect the parameters used in a functional to the LEC of the interaction, this would render the DFT completely non-empirical. This would
also be at variance with the strategy used nowadays to construct a DFT. Indeed, most currently used DFTs in cold atoms or in nuclei are usually directly adjusted either on experimental datas
or on pseudo-datas obtained using one of the ab-initio methods mentioned above. Such direct fitting procedure is very powerful because it includes automatically complex many-body correlations in the DFT. It also leads in general to a very precise description of the global properties of Fermi systems. This is for instance the case in atomic nuclei were
the precision on the ground state energy is better than 2--3 $\%$ for medium mass nuclei and goes down to 0.5 \% for heavy systems (see for instance \cite{Ben05,Nav18}).
This strategy has also some drawbacks. Among them, we usually face the difficulty that some components of the functional are not really constrained by the experiments. This is for instance the case of the density dependence of the symmetry energy in nuclei that is particularly important for the physics of exotic nuclei. Another example that was pointed out recently is the failure of empirical functionals to properly describe the low density limit of neutron matter \cite{Yan16}.
One should mention that, with recent progresses in the nuclear interaction and in ab-initio many-body techniques, there is an increasing interest in
developing DFTs directly starting from a clear many-body framework. Among the recent works, we mention the Density Matrix Expansion (DME) proposed already some times ago
\cite{Neg73,Neg75} that has reached now a certain level of maturity \cite{Geb10,Sto10,Car10,Geb11,Bog11,Dyh17}.
Another clearly defined approach is to write the effective action and use the inversion method as proposed in \cite{Plu03,Bha05} (see also the recent interesting progress of Ref. \cite{Yok19}).
Still, the quantitative description of strongly interacting systems beyond the low density limit and/or beyond the Hartree-Fock approximation is a rather difficult task.

For this reason, we explore here qualitatively how a DFT can be obtained for three dimensional infinite Fermi systems where the parameters of the DFT are
directly linked to the LEC of the interaction. More specifically, we consider the simplified problem where the interaction is described by a single LEC, $a_s$, and where the interaction
strength can vary from the perturbative to the non-perturbative regime. Such physical situation was explored in different regimes using standard many-body
techniques starting from an Effective-Field-Theory approach \cite{Fur00a,Fur08a}. For instance, the low density limit was studied in Ref. \cite{Ham00}. This case is particularly highlighting
since in this case, up to third order in perturbation, the energy can be written as a simple polynomial {(and potentially polylogarithmic from fourth order)} of $\rho^{1/3}$ where $\rho$ denotes the density. The perturbative
approach breaks down when $(a_s k_F)$ increases. In this case, DFT have also been obtained using diagrammatic resummation techniques \cite{Sch05,Kai11}. Both
perturbation and resummation to obtain compact expressions of the energy in terms of $(a_s k_F)$ will be briefly discussed here. As we will see,
the brute-force resummation however generally suffers from the lack of predictive power especially close to the unitary limit. Following Ref. \cite{Sch05}, we show that,
using a procedure called hereafter phase-space average approximation, the energy can be written as a simplified functional of $(a_s k_F)$ that in addition improves the
description of strongly interacting systems. The work of Ref. \cite{Sch05} was actually the starting point of several new developments in the nuclear many-body context.
In Ref. \cite{Yan16}, guided by the simplified expression of the energy a hybrid functional was proposed where some of the parameters are directly connected to $a_s$.
Similarly, in \cite{Lac16a,Lac17a}, a non-empirical functional was proposed that could reproduce both cold atoms gases and neutron matter up to
$\rho\simeq 0.01$ fm$^{-3}$ including the effective range effect. Such new functionals were also used in Ref. \cite{Lac17a} (see Fig. 6 of this reference)  to understand
the quantitative values of parameters that are used in empirical functional like Skyrme DFT.  It was shown that the LEC are strongly renormalized due to in-medium effects. This actually was also recently shown using Brueckner-Hartree-Fock in Ref. \cite{Zha18} and was encoded in the ELYO functional through density dependent coupling constants in Ref. \cite{Gra17}. A review on the novel scientific activities
in this field can be found in Ref. \cite{Gra18} (see also the recent work \cite{Bon18} for application to finite systems including pairing).

The thermodynamical properties of strongly interacting systems was studied in Ref. \cite{Bou18} using one of the functional proposed recently. While most of the observed
properties of systems close to unitarity were reproduced very accurately,  two difficulties have been identified. The first one is that the dynamical response function in the superfluid phase
can without surprise only be achieved by introducing explicitly the pairing field in the functional. The explicit treatment of superfluidity is not the subject of the present work and we will concentrate on normal systems. A second source  of difficulty is the absence of clear prescription for the effective mass in the large $a_s$ limit.  Such effective mass and
more generally quasi-particle properties are rather standard quantities helping to understand Fermi liquids. Its knowledge are of particular importance for instance to
understand certain properties like the static response of neutron matter recently calculated with an ab-initio theory in \cite{Bur16,Bur17}. It turns out, for instance
for neutron matter, that the effective mass in neutron systems is scarcely known (see Fig. 6 of Ref. \cite{Bou18}) and has only been very recently estimated using
AFDMC in Ref. \cite{Ism19} and BHF calculation \cite{Isa16,Are16}.
For this reason, we also explore
the possibility to obtain self-energies, for which direct contact with the Fermi liquid theory can be made, as functionals of $(a_s k_F)$ in the non-perturbative regime.
In order to achieve this goal, we also use resummation techniques and extend the phase-space average technique to the self-energy.  Finally, we  briefly discuss
how such analytical form can be useful in DFT approach.

\section{DFT for dilute systems from many-body perturbation theory}

We concentrate here on systems where the only low energy constant (LEC) at play is the s-wave
scattering length $a_s$. Infinite systems composed of spin-degenerated particles of mass $m$, i.e.
a relevant situation for non-polarized neutron matter and/or spin degenerated cold atoms, are investigated. Following Ref. \cite{Ham00} and using the Effective-Field Theory approach for homogeneous dilute Fermi gas, the s-wave interaction is simply written as a zero-range interaction that identifies with a constant in momentum space:
\begin{eqnarray}
\langle \bm{k^\prime} | V_{\rm EFT}|  \bm{k} \rangle &=& C_0,   \label{eq:C0}
\end{eqnarray}
{where $\bm{k}$ and $\bm{k^\prime}$ are the relative momenta of the incoming and outgoing particles.}
The constant $C_0$ is linked to the scattering length $a_s$ through ($\hbar=1$):
\begin{eqnarray}\label{eq:def-C0-as}
C_0 = \frac{4\pi a_s}{m}.
\end{eqnarray}
{using the convention that a negative $a_s$ is attractive, so that the the s-wave scattering phase shift $\delta_s$  verifies $k\cot \delta_s = -1/a_s$.}
The model case, where the interaction is dominated by $a_s$ has been widely exploited in Fermi systems in the past
\cite{Hua57a,Hua57b,Lee57b,Abr58,Bel61} (see also \cite{Fet71a}). The interaction (\ref{eq:C0}) has a well-known ultra-violet (UV) divergence.  In the present sections, we summarize some known results for this model.
Note that, the results have been obtained with proper treatment of the  UV  divergence  using standard techniques {(in particular minimal subtraction scheme of dimensional regularization). For more details see \cite{Bol72a,Lep97a,Kap98a,Kap98b,Phi98a,Bir98a,Bed02a}.}

Hereafter, we use the two-body Hugenholtz energy diagrams convention for homogeneous dilute Fermi system based on particle-hole propagator:
\begin{equation}\label{eq:GFph}
  G(\omega,\bm{k}) = \frac{1-n(k)}{\omega - e(k) + i\eta} + \frac{n(k)}{\omega - e(k) - i\eta}.
\end{equation}
The great simplification in infinite system stems from the fact that the relevant single-particle states for the Free-Gas (FG) are
identified as plane-waves with energy $e(k) =k^2/ 2m$ and occupation numbers at zero temperature given by $n(k) =  \Theta(k_F - k)$ where $k_F$ is the Fermi momentum of the system.
The Feynman rules with the proper symmetry factor to estimate a contribution to the energy can be found in Ref. \cite{Ham00}. The diagrams contributing to the ground state energy up to third order are shown in table \ref{tab:feyn-rules-selfenergy}.
Note that second order diagrams, i.e. composed by two $C_0$-vertex (black dot), $(2a)$ and the third order diagrams  $(3a-c)$ are vanishing in this context.

We focus first here on the possibility to obtain the energy as a function of the quantity $(a_s k_F)$. The natural many-body approach if $(a_s k_F) \ll 1$ is to start from a perturbative expansion in powers of $(a_sk_F)$. We can then expand the energy as a series:
\begin{eqnarray}
\frac{E}{E_{\rm FG}}&=&1 + \frac{ E^{(1)}}{E_{\rm FG}} +  \frac{ E^{(2)} }{E_{\rm FG}}  +  \frac{ E^{(3)} }{E_{\rm FG}} + \cdots \label{eq:pert123}
\end{eqnarray}
where $E^{(1)}/E_{\rm FG}$ is linear in $(a_sk_F)$, $E^{(2)}/E_{\rm FG}$ is quadratic in $(a_sk_F)$, ... $E_{\rm FG}$ denotes here the Free-Gas (FG) energy defined as:
\begin{eqnarray}
  E_\mathrm{FG} &=& g\int \frac{d^3k}{(2\pi)^3} n(k)e(k) = \frac{3}{5}\frac{k_F^2}{2m}\rho, \label{eq:efg}
\end{eqnarray}
where the spin degeneracy $g$ is equal to $2$ for the present case.
Here we have use the fact that the Fermi momentum is linked to the density $\rho$ through:
 \begin{eqnarray}
\rho &=& g\int \frac{d^3k}{(2\pi)^3} n(k) = \frac{g}{6 \pi^2} k^3_F . \label{eq:rhokf}
\end{eqnarray}

The perturbative approach leads to the so-called Lee-Yang expansion of the energy \cite{Hua57a,Hua57b,Lee57b}.
The same expansion was obtained in Ref. \cite{Ham00} using EFT technique.
Keeping in mind that only connected diagrams contribute to the energy,  the first order term in Eq. (\ref{eq:pert123}) is given by:
\begin{eqnarray}
	E^{(1)} &=& \nonumber \\ [-0.5cm]
&&  \includegraphics{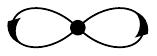}   \nonumber \\[-2em]
          && \hspace{3.75em} = E_\mathrm{FG}(g-1)\frac{10}{9\pi}(a_sk_F),
          \label{eq:E1}
\end{eqnarray}
where we recall the associated diagram.
We recognize here the Hartree-Fock (HF) contribution at leading order in $(a_sk_F)$.
The second order contribution stems from the direct and exchange terms due to the coupling between the
2 particle-2 hole (2p-2h) excited state and the uncorrelated HF ground state. It is given by:
\begin{eqnarray}
    &&\nonumber \\[-0.8em]
	  E^{(2)} &=& \nonumber \\ [-.85cm]
&& \includegraphics{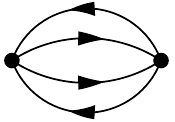}   \nonumber \\[-2.85em]
	  &&  \hspace{4.25em} = E_\mathrm{FG} (g-1)\frac{4}{21\pi^2}(11-2\ln2)(a_sk_F)^2 . \label{eq:E2} \\[-.8em] \nonumber
\end{eqnarray}
Higher order contributions can be evaluated analytically (or numerically) in a similar way. For instance,  the energy per particle at
third order has been historically obtained by Efimov and Amusia \cite{Efi65,Amu68}, Baker \cite{Bak71} and Bishop \cite{Bis73} and was more recently discussed in different works \cite{Kai11,Ste00,Pla03,Mec15a,Mec15b,Hol17a,Kai17}.
Very recently, the fourth order  has also been worked
out \cite{Wel18}. For now on, we will systematically assume that $g=2$.

\subsection{Link with DFT and shortcomings}

Since our goal is to make connection between many-body approaches starting from the bare interaction and
DFT, let us make simple early stage remarks relevant for the present discussion.
Once the energy is written in terms of powers of $k_F$, this energy can directly be interpreted as a functional of the local
density $\rho$ using the equation (\ref{eq:rhokf}).   For infinite systems, the local density is just a constant. A standard technique
to export a DFT in infinite system to finite systems is to use a Local Density Approximation (LDA) where the equation (\ref{eq:pert123}) is first
transformed into an integral over space of the energy density functional:
\begin{eqnarray}
E &=& \int d^{3} r ~{\cal E}(\bm{r}) .
\end{eqnarray}
Here,  ${\cal E}(\bm{r})$ contains the kinetic and the potential energy contributions that are both written in terms of the local density $\rho(\bm{r})$,
itself linked to a local Fermi momentum $k^3_F(\bm{r}) = 6 \pi^2 \rho(\bm{r})/g$.   This approach, that has some connections with the Thomas-Fermi approximation \cite{Rin80,Lip08}, leads to functionals of the local density consistently with the Hohenberg--Kohn theorem \cite{Hoh64}. Such direct mapping has initiated several novel ways to obtain DFT in the nuclear physics context \cite{Yan16, Gra17,Bon18} (for a review see
\cite{Gra18}). Similar strategy is now currently used in a different context in order to incorporate quantum corrections that might stabilize quantum droplets \cite{Bul02,Fer16}.

The simple strategy discussed above to obtain a DFT for many-body systems has several shortcomings:
\begin{itemize}
  \item[(i)] The perturbative approach provides a systematic and constructive approach to write the energy as a function of a polynomial of $k_F$ and/or $(a_s k_F)$ of increasing orders. It however faces the difficulty that the number of diagrams to be evaluated and/or the complexity of the integrals that appear both significantly increase from one order of the perturbation to the next order.
  \item[(ii)] Unless the expansion is made up to infinity in Eq. (\ref{eq:pert123}), the deduced energy only applies below a certain
value of the density and/or  $(a_s k_F)$.  When the s-wave scattering length becomes large, the perturbative expansion at low density is not valid anymore.
Typical examples in nature with large $a_s$ are unitary Fermi gas \cite{ZwergerBook,Chi10} or nuclear neutron matter \cite{Fet71a}.
For instance, the s-wave scattering length for neutron-neutron or proton-proton interaction is $a_s \sim -20~\mathrm{fm}$, leading to a range of validity in density $\rho \lesssim 10^{-6} ~\mathrm{fm}^{-3}$ for the perturbative expansion.
Compared to the saturation density $\rho_0 \sim 0.16 ~\mathrm{fm}^{-3}$, i.e. typical density in nuclear system, the perturbation theory is not appropriate to describe properly these systems at the relevant density scale.
  \item[(iii)] A pure LDA approximation misses some important aspects that affect the system properties.  An illustration is the effective mass $m^*$ that is standardly used in the Fermi Liquid Theory (FLT) approach.
Consistently with the expansion of the energy given by Eq. (\ref{eq:pert123}), one might also obtain an expansion of the effective mass as:
  \begin{eqnarray}
\frac{m^*}{m} &=& 1 + \left(\frac{m^*}{m}\right)^{(1)} + \left(\frac{m^*}{m}\right)^{(2)} +  \left(\frac{m^*}{m}\right)^{(3)} + \cdots \label{eq:m123}
\end{eqnarray}
In the present case where only $a_s$ is considered, a correction to the bare mass starts to appear only at second order in perturbation, leading to the so-called
Galitskii formula \cite{Gal58} (see section \ref{sec:selflow}). Again, the expansion (\ref{eq:m123}) truncated at a given order is usually restricted to the low density region.
\end{itemize}

The difficulties (i) and (ii) can eventually be   solved by treating the problem numerically in a non-perturbative framework.
With the increase of interest of systems with varying s-wave scattering length, several efficient numerical approaches have been developed. To quote some of them, we mention the Brueckner Hartree-Fock (BHF)  \cite{Dog14,Fet71a,Bru55a,Gol57a,Are16a,Isa16a},  the Self-consistent Green function (SCGF) \cite{Hau07a,Hau09a,Sop11a},
the Quantum-Monte Carlo (QMC) approach and/or Auxiliary-Field Diffusion Monte-Carlo (AFDMC) \cite{Cha04a,Gez08a,Car12a,Fri81a,Akm98a,Gez10a}, or the recently proposed approach based on Blod Diagrammatic Monte Carlo \cite{Hou12a,Hou13a,Ros18}. These direct numerical techniques, while very effective in some cases, do not
lead to an energy written as a functional of $(a_s k_F)$. Their results can still serve as pseudo-datas on which a DFT can be adjusted. This is particularly useful when properties of strongly correlated Fermi systems cannot be directly probed experimentally.

Here, however, we follow a different goal that is to have an analytical guidance to design a DFT for systems with large scattering length.
Alternatively  to a direct numerical method, the problem can be approximately solved by selecting certain classes of diagrams
and by summing up these diagrams to all orders \cite{Sch05,Kai11,Fet71a,Ste00}.  As we will
see below, this approximate treatment has also the advantage to automatically lead to analytical functions of $(a_s k_F)$.
In the present work, our strategy is to use the resummation technique as a starting point to obtain a DFT for Fermi systems
in the non-perturbative regime.   Illustrations of the resummation technique applied to the energy are given in the next section.
We then discuss, how a DFT can be deduced from it. As we will see below, considering directly the energy does not provides information on quasi-particle properties (item (iii)). For this reason, the methodology employed for the energy is extended to the self-energy in section \ref{sec:self}.

 \section{Diagrammatic resummation technique for the energy}

Motivated by systems with anomalously large scattering lengths, several strategies have been proposed
for selecting certain classes of diagrams and for providing  compact expressions of the energy in terms of $(a_s k_F)$.
In this section, we summarize some recent attempts to make resummation
of diagrams using the interaction (\ref{eq:C0}).

In 2000, Steele \cite{Ste00} laid the groundwork to describe the ground state properties of a Fermi system at low density with large scattering length based on resummation. He calculated explicitly the energy per particle up to fourth order in perturbation for ladder and ring diagrams.
{In this work, each contribution is written as an integral over the phase-space of a combination of the particle-particle, hole-hole, particle-hole (resp. hole-particle) scattering loop functions.}
Noting that the contributions of energy diagrams composed by $(n-1)$ particle-particle loops are dominant when the order $n$ of perturbation increases, he proposed to retain only these contributions and sum them up to obtain a geometric series to be integrated in the accessible phase-space.
\begin{figure}[htbp]
\hspace*{2.cm}\includegraphics{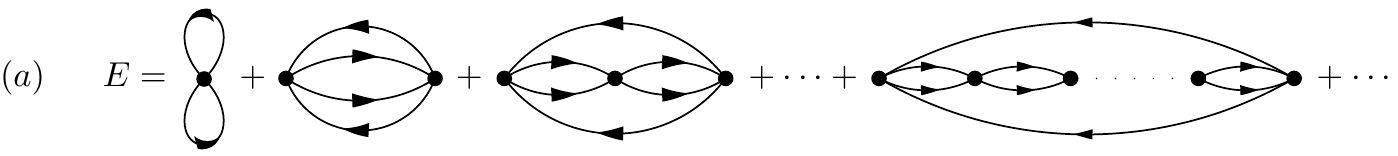}  \\
\hspace*{2.cm}\includegraphics{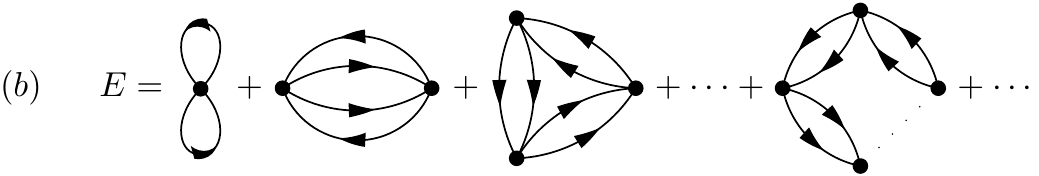}
  \caption{\label{fig:ladderSum}
  Diagrammatic representation of the energy resummation (a) Resummation of particle-particle ladder diagrams. Except for the the Hartree-Fock diagram, a line from left to right represents a particle and from right to left a hole; (b)  Resummation of combined particle-particle and hole-hole ladder diagrams. Except for the Hartree-Fock diagram, each pair of equivalent lines represent indistinguishable particles pair or holes pair.
  }
\end{figure}
The resulting energy is schematically represented in Fig. \ref{fig:ladderSum}-a as an infinite sum of diagrams.
Although rather tedious, the technique to obtain resummed expression  is rather standard starting for instance from
Green-function approach (see for instance \cite{Fet71a}). After averaging over angles {and using the minimal subtraction scheme of dimensional regularization \cite{Bol72a,Lep97a,Kap98a,Kap98b,Phi98a,Bir98a,Bed02a}}, the energy can be written
in the compact form (see for instance \cite{Sch05,Kai11,Ste00} and  \ref{app:eresum} for the definition of $s$ and $t$):
\begin{eqnarray}
  \frac{E}{E_\mathrm{FG}} = 1 + \frac{80}{\pi}\int_0^1 s^2 ds \int_0^{\sqrt{1-s^2}}  tdt ~ \frac{(a_sk_F) I(s,t)}{1  - \left( a_sk_F /\pi\right) F(s,t) } . \label{eq:GEI}
\end{eqnarray}

\begin{figure}[htbp]
\begin{center}
  	\includegraphics{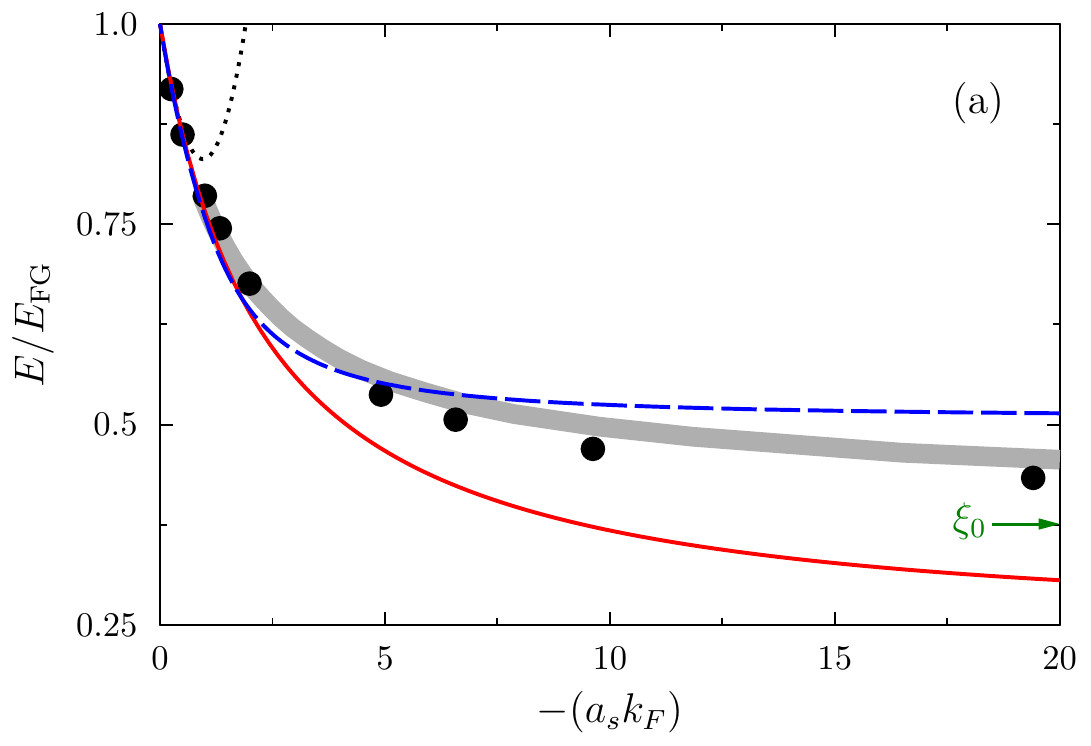} \\
  	\includegraphics{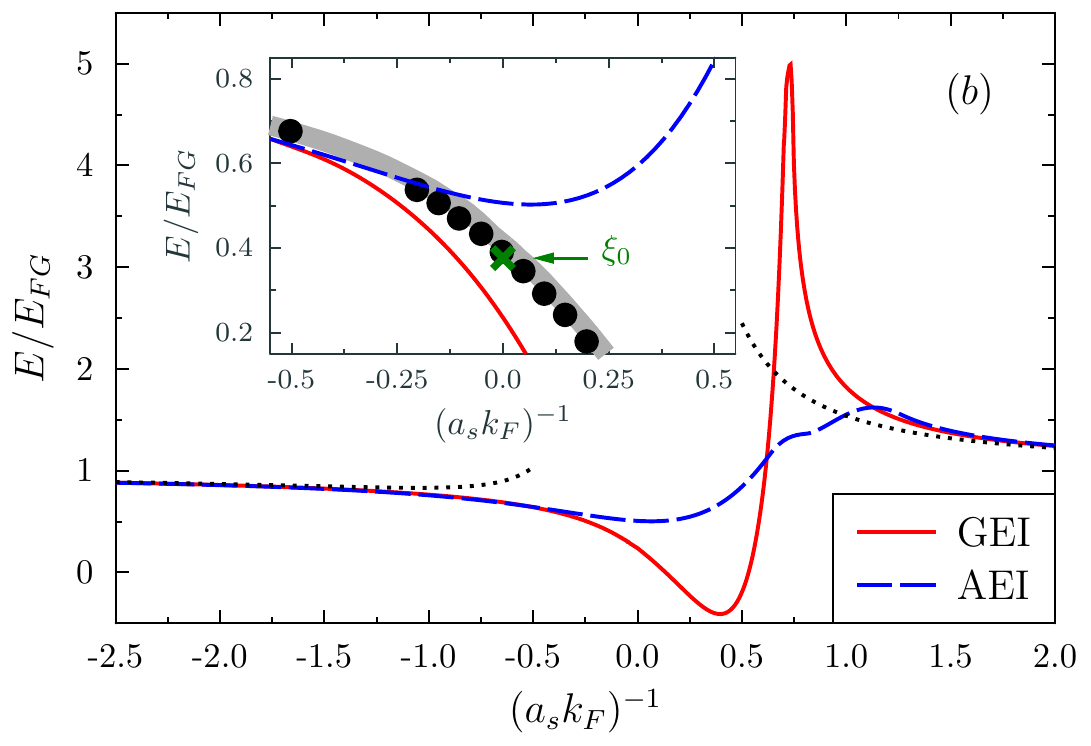}
\end{center}
  	\caption{Energy in unit of the free Fermi gas energy as function of (a) $-(a_sk_F)$ and (b) $-1/(a_s k_F)$.
	The red solid and blue dashed lines are respectively
	obtained by direct numerical integration of Eqs. (\ref{eq:GEI}) and  (\ref{eq:AEI}).
       For reference, the black dotted line correspond to the Lee-Yang formula obtained by summing up the contributions (\ref{eq:E1})
       and (\ref{eq:E2}). In both panels, the grey area indicates the result obtained by fitting the experiment \cite{Nav10} with a Pad\'e approximation while the black circles are the Diffusion Monte-Carlo
       (DMC) results
       or Ref. \cite{Car12a}. In the inset of panel (b), a focus is made near unitarity. In all panels, the arrow indicates the value of the Bertsch parameter
       $\xi_0 = 0.37$.
  }\label{fig:GEIAEI}
\end{figure}

{In the seminal work of Ref. \cite{Kai11} based on the vacuum-medium propagator consisting to rewrite the particle-hole propagator (\ref{eq:GFph}) as:
\begin{equation}
  G(\omega,\bm{k}) = \frac{1}{\omega - e(k) + i\eta} - 2\pi n(k)\delta(\omega-e(k)),
\end{equation}
a larger class of diagrams were considered by the reorganization of the many-body diagrammatic calculation (see also \cite{Kai13} for further details).
The diagrammatic representation of this resummation is shown in Fig. \ref{fig:ladderSum}-b.
Contrary to the resummation of particle-particle ladder diagram at $n^\mathrm{th}$ order by considering only particle-particle loops, this resummation takes into account diagrams composed by combined particle-particle and hole-hole loops.}
Paying particular attention to the combinatorial occurrence of each diagram, it was shown that the energy writes as
a phase-space integration of an arctangent function  \cite{Kai11}:
\begin{eqnarray}\label{eq:AEI}
		\frac{E}{E_\mathrm{FG}} = 1 + \frac{80}{\pi}\int_0^1 s^2 ds \int_0^{\sqrt{1-s^2}}  tdt ~\arctan \frac{(a_sk_F)I(s,t)}{1 - (a_sk_F/\pi)R(s,t)}.
\end{eqnarray}
In the following, results of the numerical integrations of Eqs. (\ref{eq:GEI}) and  (\ref{eq:AEI}) will be referred as
 \emph{Geometric series Exact Integration}  (GEI) and \emph{Arctangent series Exact Integration} (AEI) respectively.
Both Eqs. (\ref{eq:GEI}) and  (\ref{eq:AEI}) present several interesting features compared to a perturbative expansion.
Firstly, they could be expanded in powers of $(a_s k_F)$ and, noteworthy, their second order expansions match the Lee-Yang
  formula. Note that the Eq. (\ref{eq:AEI})  is slightly more predictive in the sense that it is valid up to third order
  in $(a_s k_F)$\footnote{Only the ladder diagram contribution term at third order is accounted for
   since ring diagrams composed by particle-hole loops (the non-vanishing diagram $(3e)$ of table \ref{tab:feyn-rules-selfenergy}) are not taken into account in the resummation.}.
   We show in Fig. \ref{fig:GEIAEI}-a a comparison of the energy obtained by integrating numerically the two resummed expressions
   as a function of $(a_s k_F)$.

One of the motivations  for the use of resummation is that, contrary to any truncation, the energy is not diverging as $|a_s k_F| \rightarrow + \infty$, i.e. in the unitary gas regime (see panel (b) of Fig. \ref{fig:GEIAEI}). This was firstly discussed in Ref. \cite{Sch05} for the GEI
  case and latter in \cite{Kai11} for the AEI case.  As noted in these works, the ratio:
  \begin{eqnarray}
\xi_0 &=& \lim_{|a_s k_F| \rightarrow + \infty} \frac{E}{E_{\rm FG}}
\end{eqnarray}
generally refereed to as the Berstch parameter, significantly depends on the class of diagram selected for resummation. In the two cases considered here, we have:
\begin{eqnarray}
\xi_{\rm GEI} \simeq 0.24 \hspace*{1.5cm} {\rm and}  \hspace*{1.5cm} \xi_{\rm AEI} \simeq 0.51 .
\end{eqnarray}
These values in both cases significantly differ from the experimentally observed value of the Bertsch parameter $\xi_0 = 0.37$ \cite{Car12a,Hau09a,Nav10,Ku12}. It should be however kept in mind that the value of $\xi_0$ corresponds to the one of a superfluid unitary gas while superfluidity is not accounted for in the present resummation. Therefore, to be consistent, one should a priori compare
with the value of the Bertsch parameter in normal systems. In \cite{Dog14}, using Brueckner Hartree-Fock approach, a value $0.507$
was found, {which is compatible with the experimental result of Ref. \cite{Ku12} giving $0.45$}. This value is actually consistent with the AEI case given by Eq. (\ref{eq:AEI}).
 Nevertheless, one obvious conclusion is that the choice
of certain diagrams significantly affects the energy behavior as $(a_s k_F)$ increases.

\subsection{Phase-space average approximation for the re-summed energy}
\label{sec:pse}

Despite the fact that the selection of diagrams influences the results of a resummation approach, the resulting expressions
of the energy in terms of $(a_s k_F)$ is an interesting steps towards a DFT for interacting systems beyond the Lee-Yang formula. Still, the deduced expressions are rather complicated especially due to the necessity to perform explicit integrations on phase-space for all
values of $k_F$ (note also that $k_F$ also appears in the definition of $s$ and $t$, see \ref{app:eresum}).  This complexity can however
be partially reduced using what we call below a Phase-Space (PS) approximation.
\begin{figure}[htbp]
\begin{center}
  	\includegraphics{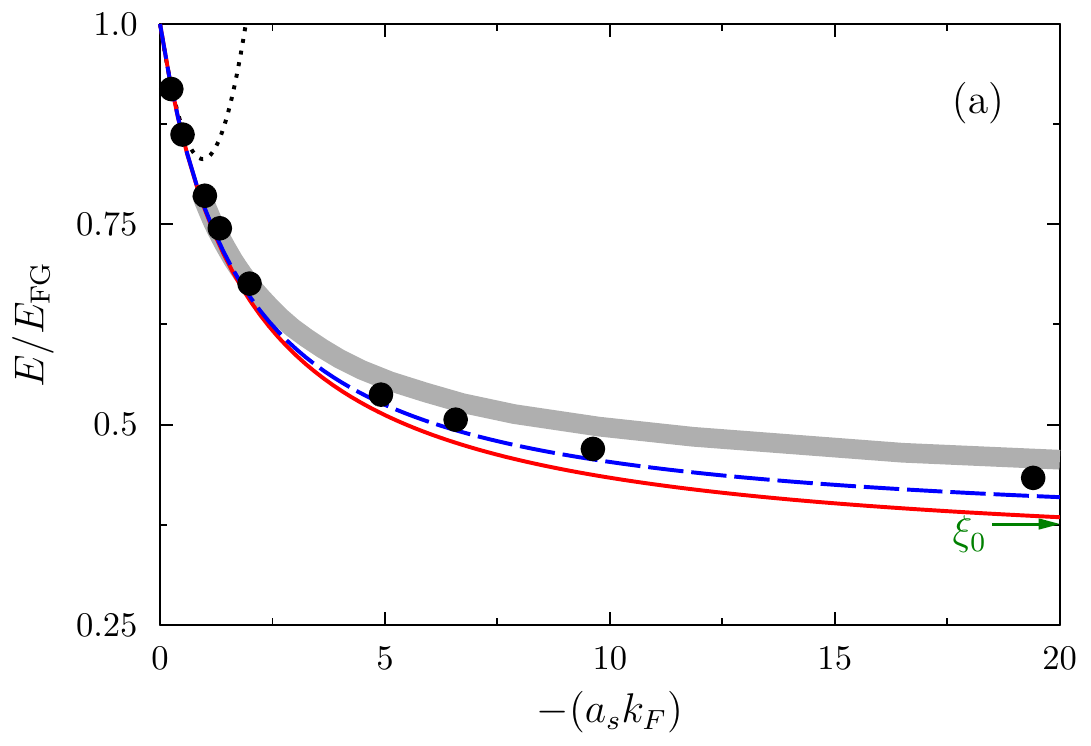} \\
	  \includegraphics{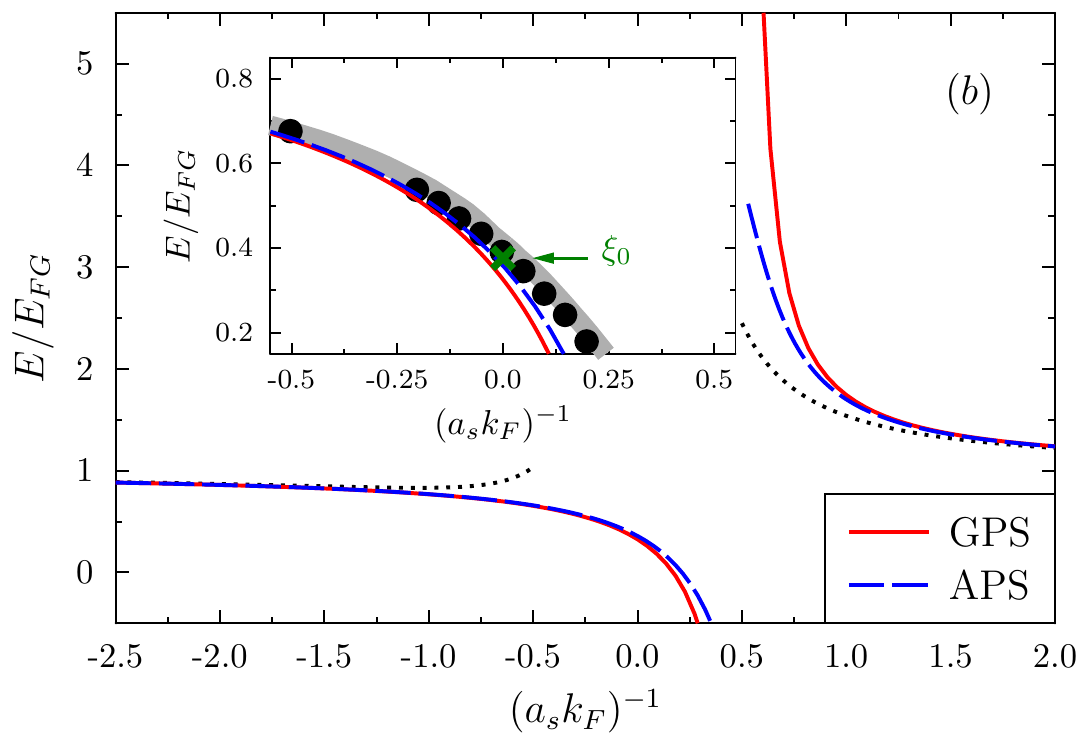}
\end{center}
  	\caption{Same as Fig. \ref{fig:GEIAEI} using the GPS (red solid line) and APS (blue dashed line) approximation.
	}
 \label{fig:GPSAPS}
\end{figure}
The PS approximation was discussed for the Geometric series case in Ref. \cite{Sch05}, it
consists simply in replacing the numerator and denominator entering in the integral respectively by their average
values integrated over the phase-space, leading in this way to a much simpler approximation. Let us introduce
the notation $\langle \langle X \rangle \rangle$ defined as:
\begin{eqnarray}
\langle \langle X \rangle \rangle \equiv \int_0^1 s^2 ds \int_0^{\sqrt{1-s^2}} tdt  X(s,t) . \nonumber
\end{eqnarray}
We see for instance that the GEI can be written as:

\begin{eqnarray}
  \frac{E}{E_\mathrm{FG}} &=& 1 + \frac{80}{\pi} \left\langle \left\langle \frac{(a_sk_F) I} {1  - \left( a_sk_F/\pi \right) F}  \right\rangle  \right\rangle \nonumber \\
  &=&  1 + \frac{80}{\pi}(a_sk_F) \langle \langle I \rangle \rangle \left\{ 1 +  \frac{(a_sk_F)}{\pi} \frac{\langle \langle I F \rangle \rangle}{\langle \langle I \rangle \rangle}
  +  \frac{(a_sk_F)^2}{\pi^2}  \frac{\langle \langle I F^2 \rangle \rangle}{\langle \langle I \rangle \rangle}  + \cdots \right\}. \nonumber \\ \label{eq:devll}
\end{eqnarray}
The PS approximation consists in replacing this expression simply assuming:
\begin{eqnarray}
  \frac{E}{E_\mathrm{FG}} &\simeq& 1 + \frac{80}{\pi}(a_sk_F) \langle \langle I \rangle \rangle \left\{ 1 +  \frac{(a_sk_F)}{\pi} \frac{\langle \langle I F \rangle \rangle}{\langle \langle I \rangle \rangle}
  +  \frac{(a_sk_F)^2}{\pi^2}  \frac{\langle \langle I F \rangle \rangle^2}{\langle \langle I \rangle \rangle^2}  + \cdots \right\} \nonumber \\
  &=& 1 + \frac{80}{\pi}  \frac{(a_sk_F) \langle \langle I \rangle \rangle} {1  - \left( a_sk_F/\pi \right) \langle \langle I F \rangle \rangle/\langle \langle I \rangle \rangle } .
\end{eqnarray}
This approximation still insures that the Lee-Yang expression is recovered up to second order in $(a_s k_F)$. Using the integrals given
in \ref{eq:integrals}, we obtain:
\begin{eqnarray}\label{eq:defGPS}
	 \frac{E}{E_\mathrm{FG}} & = & 1 + \frac{\dfrac{10}{9\pi}(a_sk_F)}{1 - \dfrac{6}{35\pi}(11-2\ln2)(a_sk_F)} . \label{eq:GPS}
\end{eqnarray}
This compact form, called hereafter Geometric Phase-Space (GPS), was introduced at several occasions in the nuclear physics and/or cold atom context \cite{Sch05,Wel18,Bak99,Hei01}.  The energy obtained in the GPS approximation is shown in Fig. \ref{fig:GPSAPS}.

  Eq. (\ref{eq:GPS}) could be interpreted as a minimal Pad\'e approximation in $(a_sk_F)$ at low
  density, the Pad\'e$[1/1]$ recently shown for instance in Fig.  2 of Ref. \cite{Wel18}. We mention that Pad\'e approximations
  Pad\'e$[k/k]$ can be obtained (see for instance Ref. \cite{Bak99}) that could reproduce the development (\ref{eq:devll}) to a given desired order in $(a_s k_F)$ for any $k$.

  One important conclusion for the present work is that the energy obtained from Eq. (\ref{eq:GPS}) largely extend the domain of density over which it
reproduces the exact Monte-Carlo result compared to the Lee-Yang formula, i.e. compared to the second order perturbation theory.
 Essentially, above $-(a_s k_F) =1$ in Fig. \ref{fig:GPSAPS},
 the Lee-Yang expression deviates significantly from the exact calculation.
 Note that the inclusion of third order perturbation theory only slightly extend  the domain of validity.  On contrary, we see in Fig. \ref{fig:GPSAPS}-a  that the resummed
 formula follows closely the exact results and therefore it could be useful to obtain a compact form for a DFT beyond the perturbative regime.
 The GPS approximation has indeed been recently used in Ref. \cite{Yan16} to obtain a nuclear EDF where some of the parameters are directly connected to the physical s-wave scattering length, contrary to the widely used Skyrme EDF \cite{Vau72}.

Although the main goal of the present work is to obtain DFT suitable beyond  the perturbative regime, we would like to mention also that the approximated form (\ref{eq:GPS})
leads to a Bertsch parameter $\xi_{\rm GPS} = 0.32$, that is closer to the one obtain at unitarity for superfluid systems  \cite{Yan16,Sch05} compared to the one obtained with direct integration. It should
be noted however that a relatively small difference in the value of $\xi_0$ leads to large deviations in the energy due to the fact that it is multiplied by the Free-Gas energy.
For this reason, it was proposed in Eq. (\ref{eq:devll}) to relax the slightly the low density constraint and adjust directly the denominator on the unitary gas in
\cite{Lac16a,Adh08}. Such strategy turns out to be be highly predictive for systems close and/or at unitarity \cite{Lac17a,Bou18}. The functional
proposed originally in \cite{Lac16a} (assuming that the effective range cancels out, $r_e=0$) will be called {\it Geometric series Unitary Limit} (GUL) in the following.
Note however, that here our primary goal is to see how far we can go beyond the perturbative regime without adjusting any parameter.

\subsection{Phase-Space approximation with Arctangent resummation [Ladder approximation]}

Using the same approximation as above, a phase-space approximation of Eq. (\ref{eq:AEI}) can be obtained
leading to the following compact expression, called hereafter {\it Arctangent Phase-Space} (APS):
\begin{eqnarray}
  \frac{E}{E_\mathrm{FG}} = 1 + \frac{16}{3\pi} \arctan \frac{\dfrac{5}{24}(a_sk_F)}{1 - \dfrac{6}{35\pi}(11-2\ln2)(a_sk_F)}.  \label{eq:APS}
\end{eqnarray}
Illustrations of the energy dependence obtained in the APS approximation are shown in Fig.  \ref{fig:GPSAPS}.  We note that the APS closely follow the GPS case at low density. This is indeed expected since both are constructed to
match the same Lee-Yang expansion for low density Fermi gas. More surprisingly, the APS turns out to be very effective up to unitarity.
It indeeds gives a Bertsch parameter equal to $\xi_{\rm APS} \simeq 0.36$ that is very close to $\xi_0$. This is an interesting finding since, contrary to the GPS case where the unitary limit can only be reproduce at the price of degrading the description
of the low density regime, in the APS case, both low density (second order expansion in $(a_s k_F)$) and unitary limit can be very reasonably
accounted for without adjusting any parameter.

For the sake of completeness and although that we do not expect to gain so much in terms of predictive power compared to the APS, we mention that similarly to the GUL case, an AUL ({\it Arctangent Unitary Limit}) approximation can be made. In that case, relaxing the constrain on the second order
term in the Lee-Yang energy and imposing the value $\xi_0$ for the Bertsch parameter, we obtain:
  \begin{eqnarray}
    \frac{E}{E_\mathrm{FG}} = 1 + \frac{16}{3\pi} \arctan \frac{\dfrac{5}{24}(a_sk_F)}{1 - (a_sk_F) C}  \label{eq:AUL}
  \end{eqnarray}
 where $C =  \frac{5}{24}\tan^{-1}\left[\frac{3\pi}{16}(1-\xi_0)\right]$.

 The four approximations [GSP, APS, GUL, AUL] introduced here are rather simple functions of $k_F$ compared to the original GEI, AEI
 integral equations and therefore provide much simpler functionals of the density $\rho$.
 One should mention a drawback of the phase-space approximation (see Fig. \ref{fig:GPSAPS}).
 By using phase-space average in the denominator, one restrict the
 value of $(a_s k_F)$ that could be used. Indeed, while the integrated GEI and/or AEI integral forms can be applied from negative to positive
 values of $a_s$ around unitarity, this is not the case for the phase-space expressions where a pole appears for a certain positive value of $a_s$.
 From now on, we will only consider the case where $a_s$ is negative that is also the relevant situation for neutron matter.

In summary, we have shown here that several functionals can be obtained that reproduces quite well the energy of Fermi gases at unitarity.
We would like to mention that the value $\xi_0=0.37$ is the admitted value of {\it superfluid} unitary gas. It might then be surprising to reproduce this value
with a functional originally motivated by the diagrammatic expansions of Ref. \cite{Kai11,Kai13} where superfluidity is not treated. It is however
important to keep in mind that whatever is the motivation/strategy to produce a DFT, the only final criteria is the ability of the functional to  accurately describe the ground-state energy of the system at various densities. This is actually the only purpose of a DFT constructed in the spirit of the original work of Hohenberg and Kohn who have shown the existence of a functional able to reproduce the exact energy consistently with the exact
local one-body density.

In the following, however we would like to consider directly the self-energy that is a priori clearly beyond the scope of a DFT approach. In this case, the discussion made above
for the energy cannot be made and superfluidity should be explicitly introduced to describe superfluid systems. Such a treatment is beyond the scope of the present work and
in the rest of the article, we will concentrate on non-superfluid systems.

\section{Diagrammatic resummation for the self-energy}
\label{sec:self}

Scientists working on the description of finite systems using DFT also often use these DFTs to get much more information
than simply the energy of the system. Going from a Hohenberg--Kohn \cite{Hoh64} to a Kohn--Sham \cite{Koh65} framework, a Slater determinant and/or more generally quasi-particle vacuum to treat superfluidity is first introduced to construct the normal and anomalous densities.
In the nuclear physics context and although strong debates exist on the possibility to interpret physically the single-particle and/or quasi-particle properties (see for instance \cite{Lac10}), the single-particle shell evolution is standardly considered as a relevant output of the nuclear DFT \cite{Mey93, Ben03}.  It is also standard to make connection with Fermi-Liquid theory \cite{Bay08}. As noted in Ref. \cite{Bou18},
functionals that do not explicitly use the concept of effective masses and/or pairing gap will have difficulties to reproduce the
excitation properties in strongly interacting Fermi systems. This is one of the main motivation of the present  work. Starting from a many-body
diagrammatic approach based on Green functions, a natural way to make connection with the Fermi liquid theory is to use the concept of
self-energy \cite{Fet71a}. In the present section, we explore the possibility to start from a well-defined many-body framework based
on selection of diagrams followed by resummation and see if the phase-space average approach used for the energy in previous section
can be exported to the self-energy. We will use as a starting point the resummation performed in Ref. \cite{Kai13} that is consistent with the
GEI or AEI energy depending on the selected diagrams.  In Ref. \cite{Kai13}, ``only'' systems in the normal phase are considered.
Similar semi-analytical developments in the superfluid phase would require to use the Gorkov-Green approach instead of the standard Green function framework. As far as we know, although this would be a very useful while most probably extremely involved problem, this has not been achieved yet. For this reason, we will only consider normal systems here.

\subsection{Self-energy for Fermi systems at low density}
\label{sec:selflow}

We restrict the present discussion to the on-shell approximation of the self-energy, so that the self-energy
$\Sigma(k)$, {also called complex-valued single-particle potential}, is a function solely of the single-particle momentum $k$.
Self-energy are slightly more complex to obtain in the sense that more diagrams contributes to the self-energy compared to the energy. This is illustrated in table \ref{tab:feyn-rules-selfenergy} where we show the important diagrams in both cases for a contact interaction up to third order. By closing the legs, we recover the diagrams that have been estimated to obtain $E^{(1)}$ and  $E^{(2)}$ previously.

\begin{table}
\begin{center}
\begin{tabular}{p{0.4cm}c||p{0.4cm}c}
  &  \bf Self-energy & & \bf Energy   \\  &  \bf Diagrams    & & \bf Diagrams \\ \hline\hline &&& \\[-0.75em]
    $(1\alpha)$ & \includegraphics{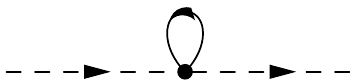} & $(1a)$ & \raisebox{-.3\height}{\includegraphics{E1a.pdf}} \\[0.5em] \hline &&& \\[-0.75em]
    $(2\alpha)$ & \includegraphics{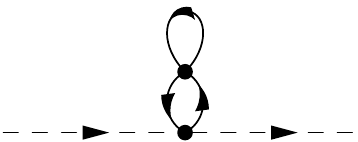} & $(2a)$ & \raisebox{-.3\height}{\includegraphics{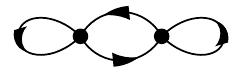}} \\[0.5em]
    $(2\beta)$  & \includegraphics{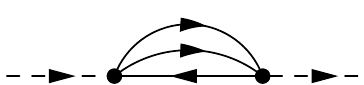} & $(2b)$ & \raisebox{-.4\height}{\includegraphics{E2b.pdf}} \\[1em]
	  \hline &&& \\[-0.75em]
    $(3\alpha)$               & \includegraphics{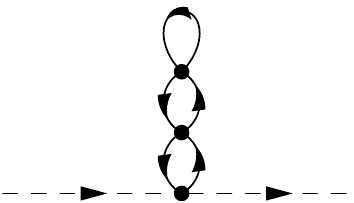}   & $(3a)$ & \raisebox{-.4\height}{\includegraphics{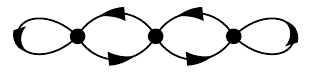}}  \\[1em]
    $(3\beta)$                & \includegraphics{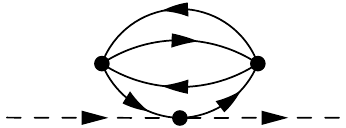}   &        &   \\
    $(3\beta^\prime)$         & \includegraphics{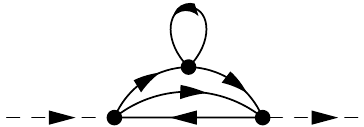}  & $(3b)$ & \raisebox{-.4\height}{\includegraphics{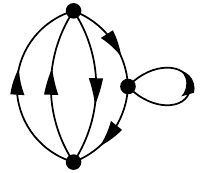}}  \\[-1.25em]
    $(3\beta^{\prime\prime})$ & \includegraphics{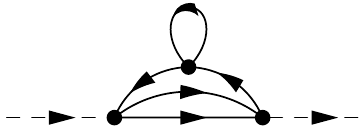} &        &   \\[1em]
    $(3\gamma)$               & \includegraphics{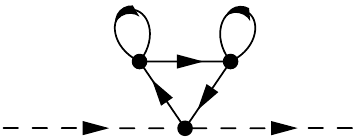}   & $(3c)$ & \raisebox{-.4\height}{\includegraphics{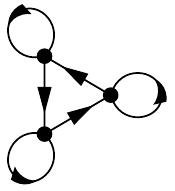}}  \\
    $(3\delta)$               & \includegraphics{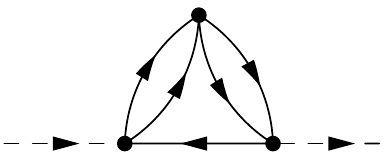}   & $(3d)$ &  \raisebox{-.4\height}{\includegraphics{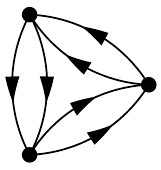}} \\
    $(3\epsilon)$             & \includegraphics{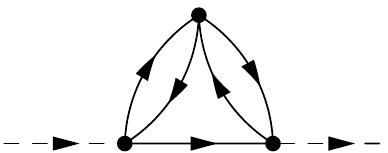}   & $(3e)$ & \raisebox{-.4\height}{\includegraphics{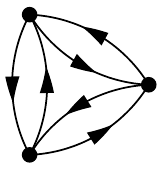}}
    \\[-0.75em]  &&&  \\  \hline \hline
\end{tabular}
\end{center}
\caption{\label{tab:feyn-rules-selfenergy}
Diagrams obtained using the Feynman rules for the self-energy contribution up to third order \cite{Pla03}.The right column corresponds to the associated energy diagrams obtained by closing the legs (dashed lines).}
\end{table}

Using the perturbative approach, the proper self-energy relevant for dilute systems
can also be expanded as:
\begin{eqnarray}
\Sigma(k) &=&  \Sigma^{(1)}(k) +   \Sigma^{(2)}(k) +  \Sigma^{(3)}(k) +\cdots\label{eq:sigseries} \\
 \nonumber \\
&=&  \nonumber \\ [-1cm]
&& \includegraphics[clip,trim=0.55cm 0cm 0.55cm 0cm]{S1a.pdf}
 + \includegraphics[clip,trim=0.2cm 0cm 0.2cm 0cm]{S2b.pdf} + \cdots
\end{eqnarray}
The first two terms are well-known and are respectively given by:
\begin{eqnarray}
	\dfrac{\Sigma^{(1)} (k)}{ \mu_{FG}}&=&\frac{4}{3\pi}  (a_sk_F),
~~~{\rm and} ~~~
	\dfrac{ \Sigma^{(2)}(k)}{ \mu_{FG}}
  = (a_sk_F)^2\Big[\Phi_2({p}) + {\rm i} \Omega_2(p) \Big].  \label{eq:self12}
\end{eqnarray}
Here, we have  defined $ \mu_{FG} \equiv k^2_F / (2m)$ and we introduced the reduced momentum ${p} = {k}/k_F$.
The two functions $\Phi_2(p)$ and $\Omega_2(p)$ related respectively to the real and imaginary parts of the self-energy
were first derived by Galitskii \cite{Gal58} (Eqs. (34) and (35) of \cite{Gal58}). For the sake of completeness, they are also recalled in \ref{app:self}.
The third order contribution to the self-energy (diagrams ($3\alpha$ -- $3 \epsilon$) in table \ref{tab:feyn-rules-selfenergy})
was studied within EFT in Ref. \cite{Pla03}.

In the following, we will introduce the notation $U$ and $W$ for the real and imaginary parts of the self-energy.
These quantities enters respectively into  the single-particle (sp) energy $\varepsilon(k)$
and the lifetime $\gamma(k)$ of the quasi-particle (qp) respectively. We will essentially focus here on the real part of the self-energy.
The quasi-particle properties can be obtained from the behavior of the self-energy close to $k=k_F$ ($p=1$) \cite{Lan57a}.
For instance, the chemical potential $\mu$ and the effective mass $m^*$ are respectively linked to the value of $U(k)$ and its derivative at $k=k_F$:
\begin{eqnarray}\label{eq:def-chemicalpot}
	\mu &=& \frac{k_F^2}{2m} + U(k_F),\hspace*{2.cm} \frac{m}{m^*} = 1 + \left. \frac{m}{k_F}\frac{\partial U(k) }{\partial k}\right|_{k=k_F}. \label{eq:mumstar}
\end{eqnarray}
Starting from the expansion (\ref{eq:sigseries}) and making a Taylor expansion of each terms around $k = k_F$ (or $p = 1$), one obtains
a systematic approach to compute quasi-particle properties in powers of $(a_s k_F)$. For instance using the Taylor expansion:
\begin{eqnarray}
\label{eq:re-galitskii-LO}
	\Phi_2(p)   &=& \frac{4}{15\pi^2}(11-2\ln2)-\frac{16}{15\pi^2}(7\ln2-1)(p-1) + \cdots
\end{eqnarray}
leads to the following expression of the chemical potential and effective mass up to second order in $(a_s k_F)$:
\begin{eqnarray}
\left\{
\begin{array}{l}
\displaystyle \frac{\mu}{\mu_{FG}} = 1 + \frac{4}{3\pi}(a_sk_F) + \frac{4}{15\pi^2}(11-2\ln2)(a_sk_F)^2 +  \cdots    \\
\\
\displaystyle  \frac{m^*}{m} = 1 + \frac{8}{15\pi^2}(7\ln2-1)(a_sk_F)^2 + \cdots
\end{array}
\right.  \label{eq:mmstarNLO}
\end{eqnarray}
These expressions are well known results also discussed in Ref. \cite{Gal58}. In particular, the latter
equation is often refereed as the Galitskii mass formula.

\subsection{Self-energy resummation}

The perturbative approach to the self-energy
is valid for weakly interacting systems in the low density regime. For this reason, as for the energy,
some attempts to provide resummed expressions have been made in the past directly on the self-energy.
{Our starting point is the resummation approach proposed in Ref. \cite{Kai13} for the self-energy identified as the first functional derivative according to the occupation number of the GEI/AEI resummed energy so that the diagram selection rules are consistent with Ref. \cite{Kai11}}. This consistency will be extremely useful when we will perform
the phase-space average on the self-energy.

\begin{figure}[htbp]
\begin{center}
\includegraphics{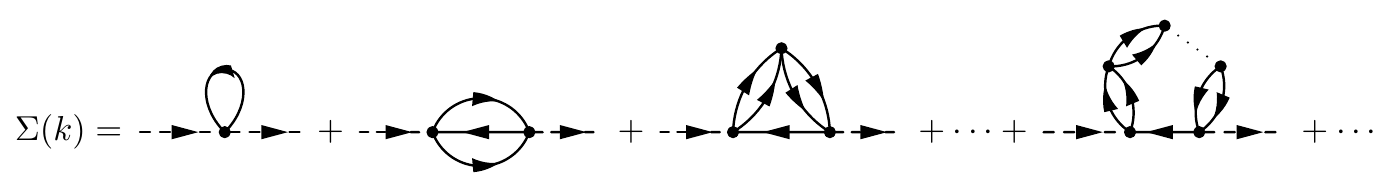}
\caption{
  Schematic diagrammatic representation of the self-energy resummation
 equivalent to the one presented in Fig. \ref{fig:ladderSum}-b.
   }\label{fig:selfresum}
\end{center}
\end{figure}

Selecting the diagrams in table \ref{tab:feyn-rules-selfenergy} that give those entering in the energy resummation  of Fig. \ref{fig:ladderSum}, a compact expression was obtained for the self-energy in  \cite{Kai13}.  Such resummation can be schematically represented as in Fig. \ref{fig:selfresum}.

\begin{figure}\centering
  \includegraphics{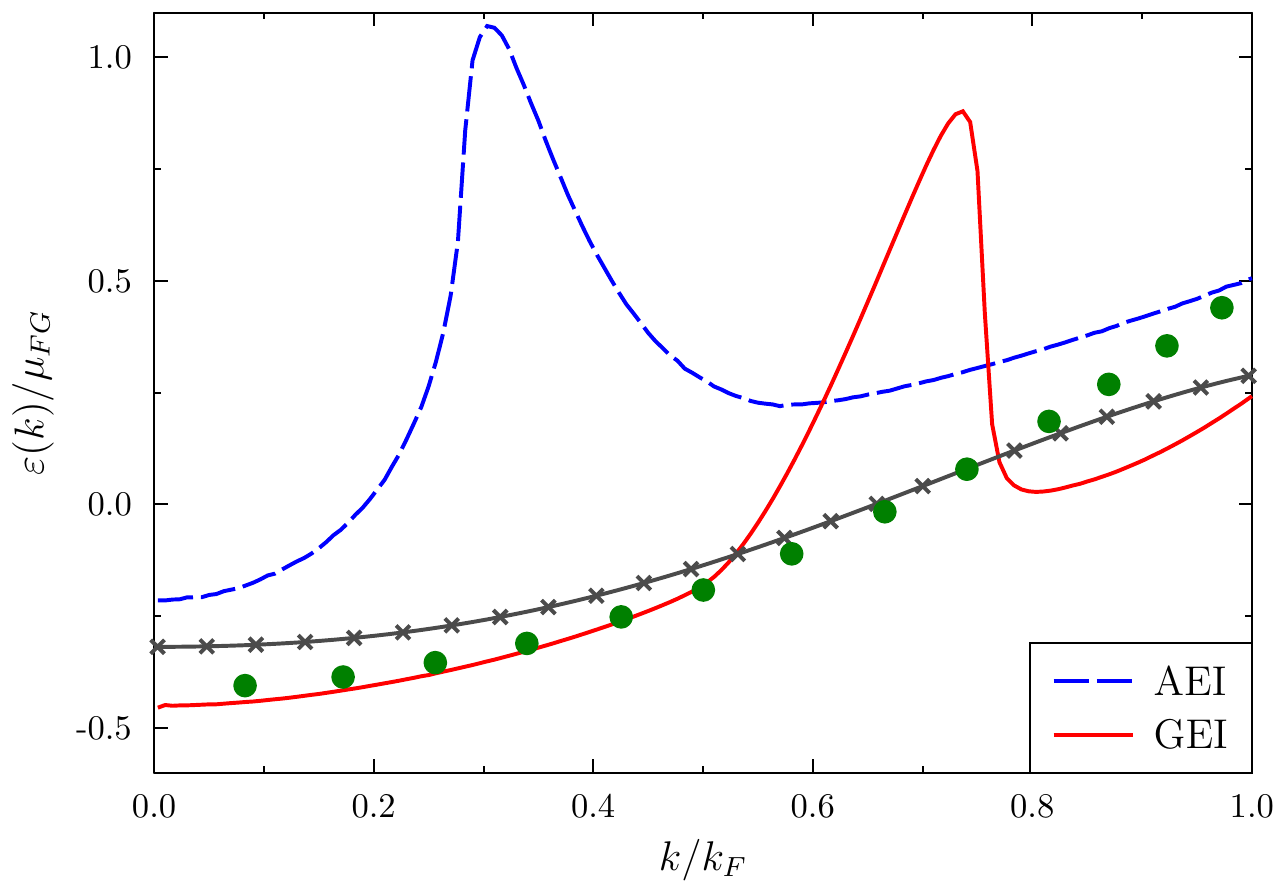}
  \caption{Single-particle energy as a function of $k/k_F$ obtained at strict unitarity by direct integration of Eq. (\ref{eq:SEintPS})
  with $pp$ ladders only [red solid line] or both $pp$ and $hh$ ladders [blue dashed line].
 For comparison, the green circles correspond to the Brueckner Hartree-Fock calculation obtained without pairing effect \cite{Dog14}
 and the black crosses to the best fit of the experimental results obtained  \cite{Ste2008a}.}
  \label{fig:GEIAEIunit}
\end{figure}

We consider here the two cases where either both particle-particle and hole-hole ladder diagrams are considered or only
particle-particle ladder diagrams are used for the resummation. These two cases are respectively consistent with the AEI and GEI
resummation for the energy. We refer to \cite{Kai13} for technical details  (see  also \cite{Bou19} for a complete discussion).
After resummation and {angle averaging approximation}, the self-energy can be written as (again with the convention $p=k/k_F$):
\begin{eqnarray}
  \frac{\Sigma (k)}{\mu_{FG}} & = &
  \Theta(k_F-k)\int_0^{1}s^2 ds \int_0^{\sqrt{1-s^2}} tdt ~ \mathcal{S}(s,t,p)
  \nonumber \\ &+&
  \Theta(k-k_F)\int_0^{(1+p)/2}s^2 ds \int_0^{(1+p)/2} tdt ~ \mathcal{S}^\prime(s,t,p).
   \label{eq:SEintPS}
\end{eqnarray}
Here $\mathcal{S}$ and $\mathcal{S}^\prime$ take different forms depending on the type of diagrams that are used.
To avoid confusion between these two cases we will use the following convention: $\Sigma (k) = U(k)+{\rm i} W(k)$
for the self-energy obtained  from the resummation of combined $pp$ and $hh$ ladder diagrams
and $\Sigma_{pp}(k) = U_{pp}(k)+{\rm i}  W_{pp}(k)$ for the one where only $pp$ ladder diagrams are used. Similarly to
the self-energy that could be separated into a real and imaginary part, we can decompose ${\cal S}$ and ${\cal S}'$ as:
\begin{eqnarray}
   \mathcal{S}^{(\prime)} & = & \mathcal{U}^{(\prime)} + {\rm i} \mathcal{W}^{(\prime)}~~~(pp~{\rm and}~hh~{\rm ladder~diagrams}),
   \nonumber \\
   \mathcal{S}^{(\prime)}_{pp}  &=&  \mathcal{U}^{(\prime)}_{pp} + {\rm i} \mathcal{W}^{(\prime)}_{pp}
~~~(pp~{\rm ladder~diagrams ~only}).
   \nonumber
\end{eqnarray}
The expressions of these functions are given in \ref{app:selfresum}. The resummed self-energies of Ref. \cite{Kai13}
have a number  of interesting properties. First, in the low density regime, the first and second order self-energies given by Eq (\ref{eq:self12}) are properly recovered
  by construction.  Another interesting feature of the expression given in \ref{app:self}, is that the self-energies do converge also to a finite result as $|a_s k_F| \rightarrow + \infty$
  for all $k$.  This is illustrated in figure \ref{fig:GEIAEIunit} where the results of direct integration of Eq. (\ref{eq:SEintPS}) are shown as a function of $k$ at unitarity. {Note that for the multidimensional numerical integration of the equations, we used the Vegas method implemented in the Cuba library \cite{Hah05}}. The single-particle energies defined as:
\begin{equation}\label{eq:epsilon-def}
  \varepsilon(k) = \frac{k^2}{2m} + U(k),
\end{equation}
and obtained in Fig. \ref{fig:GEIAEIunit} coincides with those reported in Ref. \cite{Kai11} with marked bumps. These bumps seems unphysical not only because
they significantly differ from the BHF calculations of Ref. \cite{Dog14} but also due to the presence of single-particle energies above
the Fermi energy for $k<k_F$. We would like to mention that there is no reason that  the deduced self-energy is predictive at large scattering length due to the neglected diagrams.

\begin{figure}[htbp]
\begin{center}
\includegraphics{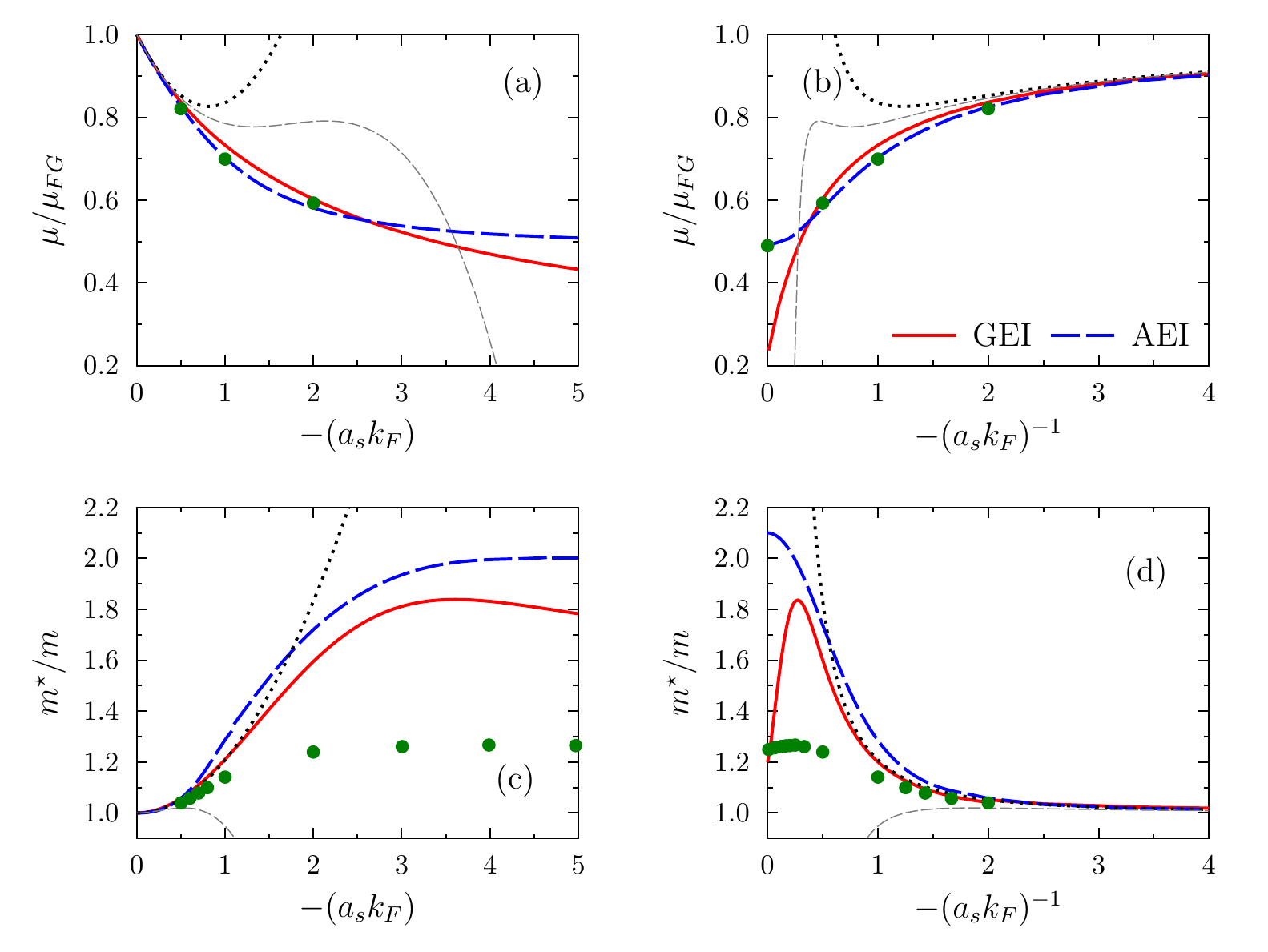}
  \caption{Chemical potential and effective mass as a function of $(a_s k_F)$ [panels (a) and (c)] or $-(a_s k_F)^{-1}$
[panels (b) and (d)] obtained with the  GEI (red solid line) and AEI
(blue dashed line) approximations. Result of the second order (Galitskii formula) and third order expansion \cite{Pla03}
in $(a_s k_F)$ are shown with black dotted and thin gray dashed lines respectively. The green circles correspond to the result of the BHF calculations of Ref.  \cite{Dog14}.}
\label{fig:mumstarkaiser}
\end{center}
\end{figure}

Starting from the different expressions, one can deduce from it, Eq. (\ref{eq:mumstar}), the quasi-particle properties by direct numerical integration. We show in Fig. \ref{fig:mumstarkaiser} the evolution of $\mu$ and $m^*/m$  as a function of $(a_s k_F)$ obtained in the two types of resummation considered here.
We see that both AEI and GEI approximation significantly extend the domain of validity compared to the perturbative theory and the corresponding chemical potential extracted from them are both rather close to the
BHF result up to $|a_s k_F| \simeq 2-3$.
For the AEI case, it is remarkable to see, especially having in mind the strange behavior of Fig. \ref{fig:GEIAEIunit},  that the chemical potential extracted from the AEI case perfectly matches the BHF calculation for all regime of $(a_s k_F)$. We see however (panels (c) and (d) of Fig. \ref{fig:mumstarkaiser}) that the comparison is in general less favorable
for the effective mass. Both approximations overpredict the effective mass compared to the BHF calculation for $|a_s k_F|> 1$,
even if the agreement is slightly better than the second or third order perturbation theory.
We also observe that the effective mass, as well as the chemical potential, obtained with each other strongly depend on the selected diagrams.

A last important remark for the discussion below is that the expression of the re-summed self-energies obtained either from
$pp$ ladders or combined $pp$ and $hh$ ladders resummations are {\it consistent} with the GEI and AEI approximations given by
Eqs. (\ref{eq:GEI}) and (\ref{eq:AEI}) respectively. Consistent means here that they respect the Hugenholtz-van-Hove (HvH) theorem \cite{Hug58a}.
This  theorem (at zero temperature) states that the single-particle energy, given by Eq. (\ref{eq:epsilon-def}), evaluated at the Fermi surface ($k=k_F$) is equal to the chemical potential of the systems.
Using the thermodynamical relation between the chemical potential and the ground state energy, $\mu = \partial E / \partial N |_{V}$, where $V$ is the unit volume, the HvH theorem leads to:
\begin{eqnarray}\label{eq:general-HvHthm}
  \frac{\mu}{\mu_{FG}} = \frac{E}{E_{FG}} + \frac{k_F}{5}\frac{\partial E/E_{FG}}{\partial k_F}
                       =  1 + \frac{{\rm Re} \left\{ \Sigma(k_F)\right\}}{\mu_{FG}}.
\end{eqnarray}
This equation gives a strong constraint between the energy and single-particle potential at $k=k_F$.

We have seen here that the resummation of diagrams is only a semi-success to predict quasi-particle properties. More specifically, the effective mass
deviates rather rapidly from the expected result as $|a_s k_F|$ increases.  Although their predictive power is limited, the AEI and GEI
approximation can serve as a guidance to provide simplified expressions of the self-energy that will be useful
latter on the DFT context. We discuss below two approaches to obtain compact expression of the single-particle potential.

\subsection{Partial phase-space average for the self-energy}
As we have seen in section \ref{sec:pse}, the phase-space average, by avoiding the estimates of rather complex integrals,
automatically led to simplified expressions for the resummed energy that turned out to be rather useful in practice \cite{Yan16,Lac16a,Bou18}.
The goal here is to develop an equivalent method directly at the self-energy level.

The first difference compared to the energy is that the phase-space average should not be made on all variables because
the self-energy depends on $k$ (or $p$).  In the following, starting from expression (\ref{eq:SEintPS}), for any function $X$ that depends on the variable $(s,t,p)$, we will introduce the two averages:
\begin{eqnarray}\label{eq:def-PSaverage-notation}
  \langle X \rangle^<_p  &\equiv&
  \int_0^{1}s^2 ds \int_0^{\sqrt{1-s^2}} \hspace{-15pt}tdt ~X(s,t,p),  \label{eq:pps1} \\
  \langle X \rangle^>_p  &\equiv& \int_0^{(1+p)/2}s^2 ds \int_0^{(1+p)/2} tdt X(s,t,p), \label{eq:pps2}
\end{eqnarray}
that denotes partial phase-space (PPS) average at fixed value of $p$ respectively relevant for $p<1$ and $p>1$.
With these notations, Eq. (\ref{eq:SEintPS}) writes:
\begin{eqnarray}
  \frac{\Sigma (k)}{\mu_{FG}} & = &
  \Theta(k_F-k)  \langle  \mathcal{S}  \rangle^<_p  +    \Theta(k-k_F)  \langle  \mathcal{S'}  \rangle^>_p .  \nonumber
  \nonumber \end{eqnarray}
Expressions of the PPS for some functions are given in \ref{app:pps}.

Starting from these expressions, our goal is to provide for the self-energy a phase-space approximation similar to the one for the energy given in section \ref{sec:pse}.
Following the strategy we used previously, we will impose the approximate form to fulfill specific constraints:
\begin{itemize}
  \item[(i)] {\bf Low density limit:}  We will always impose that the self-energy matches the exact self-energy in the low density limit
  up to a certain order in $(a_s k_F)$.
  \item[(ii)]  {\bf Large $\bm{(a_s k_F)}$ limit:} We also seek for expressions that do not diverge in the limit $|a_s k_F| \rightarrow + \infty$.
  \item[(iii)] {\bf Consistency with the HvH theorem:} while it is not a priori absolutely necessary, in some cases, we will in addition
  impose that the self-energy we obtained should be consistent with either the GPS or the APS energy. Again,
  consistency means here that the considered
  self-energy and the energy obtained through phase-space average leads to the same chemical potential using Eq. (\ref{eq:general-HvHthm}). Note that the latter condition is more
  constraining than the condition (i) and (ii). In particular, since the energy already has the constraints (i) and (ii), they will be automatically fulfilled when (iii) is explicitly imposed.
\end{itemize}

In the following, we present two strategies to get approximate self-energies using PPS. In the first strategy called hereafter {\it simple PPS} approximation, we only impose conditions (i) and (ii). While in the second strategy, that would be called {\it consistent PPS}, the form of the energy deduced will also be constrained to one of the PS approximation discussed in section \ref{sec:pse} by requiring the HvH theorem to hold.

\subsection{Simple PPS approximation for the self-energy}

We illustrate here a first simple strategy we can follow to impose (i) and (ii) avoiding complicated integrals. Our starting
point is Eq. (\ref{eq:SEintPS}).  As an illustration, we consider the specific case of the $pp$ diagrams resummation
and concentrate on the real part $U_{pp}$ of the self-energy with $p<1$.  Starting from the expressions given in \ref{app:self}, a direct
expansion of the denominator in $(a_s k_F)$ gives:
\begin{eqnarray}
\label{eq:expand-Upp}
  \frac{U_{pp}(k)}{\mu_{FG}} & = & \sum_{n=1}^\infty (a_sk_F)^n \Phi_n(p),
\end{eqnarray}
where the $\Phi_n(p)$ are given in terms of a PPS in the appendix by Equation (\ref{eq:ppsphin}).
For instance $\Phi_1(p) = 4/3\pi$ while the expression of $\Phi_2(p)$ is given by Eq. (\ref{eq:re-galitskii}).
In the following, we will often use the second order approximation:
\begin{eqnarray}
U_{pp}(k)/\mu_{FG} = 4/3\pi(a_sk_F) + \Phi_2(p)(a_sk_F)^2 + \mathcal{O}(a_sk_F)^3 \label{eq:Usecond}
\end{eqnarray}
as a reference for the low density limit.
Let us assume that we seek for an approximate form of the self-energy that matches the expansion  (\ref{eq:expand-Upp})
up to second order in $(a_s k_F)$ while being non-divergent at large value of $a_s$.  Guided by the approximation made
for the energy, one might simply use a  Pad\'e$[1/1]$ approximate form:
\begin{eqnarray}
\frac{U_{pp}(k)}{\mu_{FG}} & \simeq & \frac{4}{3\pi}\frac{(a_s k_F) }{1 - (a_s k_F) \dfrac{3 \pi}{4} \Phi_2(p)}.\label{eq:simplePPSGPS}
\end{eqnarray}
Note that, guided by the  APS expression, a similar expression can be obtained that approximate the AEI.
From these expressions
and using the expansion of $\Phi_2(p)$ given by eq.  (\ref{eq:re-galitskii-LO}), we immediately see that the low density limit
of the chemical potential and effective mass given by Eq. (\ref{eq:mmstarNLO}) are recovered. In addition starting from the expression
of $\mu (k_F)= \mu_{FG} + U_{pp}(k_F)$, the corresponding form of the energy in infinite systems can be obtained simply using the relationship:
\begin{eqnarray}
\frac{E}{N} & = &  \frac{1}{\rho} \frac{g }{2 \pi^2} \int_0^{k_F} {k^2_F} \mu (k_F) dk_F , \label{eq:emu}
\end{eqnarray}
that can easily be obtained from the definition of $\mu$ as a partial derivative of the energy with respect to the particle number.

\begin{figure}[htbp]
\begin{center}
  \includegraphics{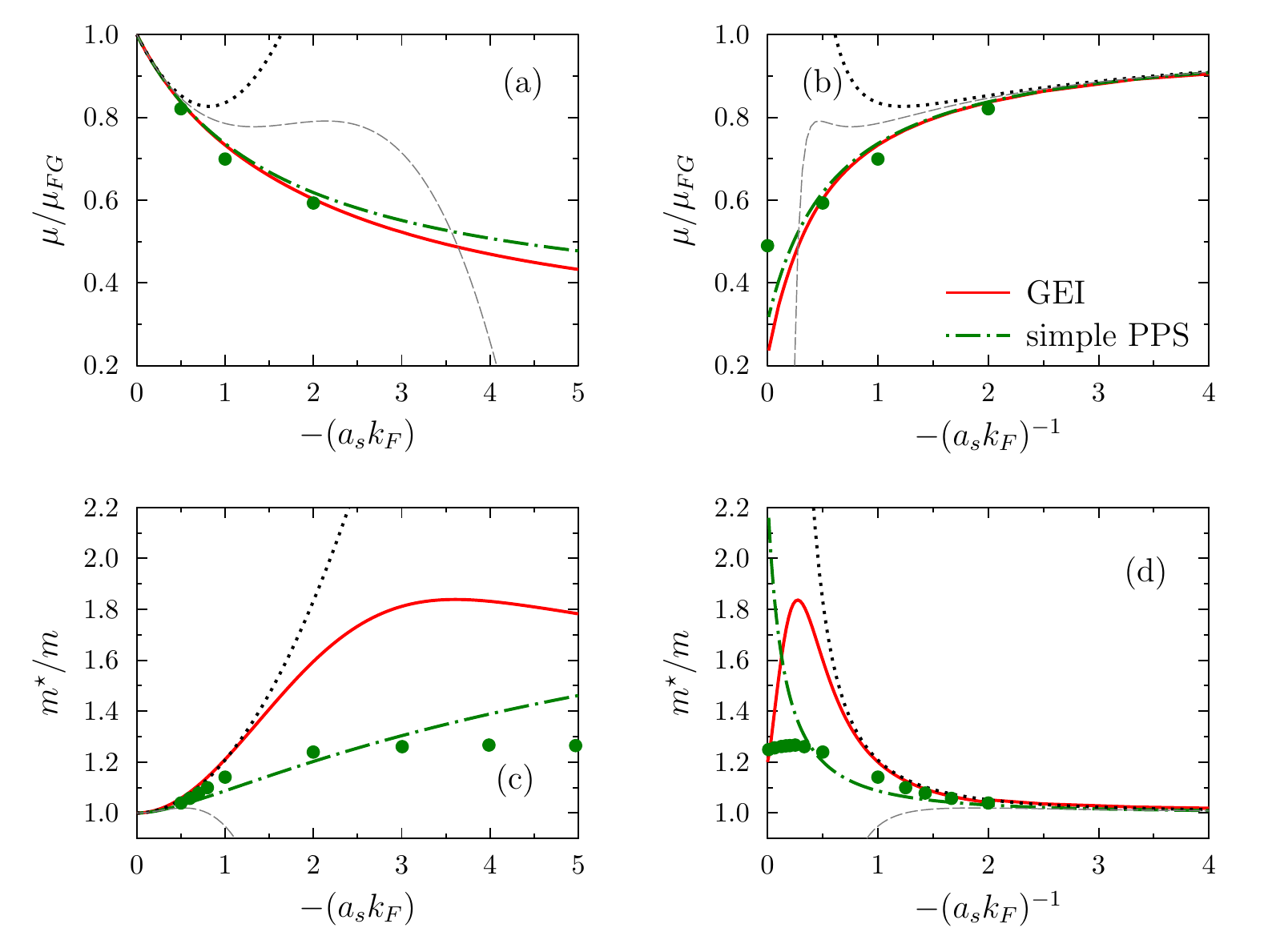}
\caption{Comparison of the chemical potential [panels (a) and (c)] and effective mass [panels (b) and (d)]
obtained with the simple PPS approximation given by Eq.  (\ref{eq:simplePPSGPS}) (green dot-long dashed line)
and the original GEI result (red solid line). For comparison, the BHF results \cite{Dog14} are also shown (green circles). }
\label{fig:simplemumstarPPSas}
\end{center}
\end{figure}

We compare in Fig. \ref{fig:simplemumstarPPSas} the dependences of the chemical potential and effective mass obtained with the simple GPS
approximation for the single-particle potential respectively as a function of $-(a_s k_F)$ or $-(a_s k_F)^{-1}$.
By construction, contrary to an approximation where the self-energy is truncated at a given order in $(a_s k_F)$,
simple GPS leads both to a smooth and converging behavior of these quantities up to infinite values of $a_s$.
These approximated self-energies also reproduce correctly the low density limit. We see in particular that the chemical potential
obtained with the simple PPS approximation follows closely the one of the original GEI result. We note also that the effective mass is more affected by the phase-space approximation. It turns out to be slightly
lower compared to the GEI case and closer qualitatively to the BHF results up to $-(a_s k_F)\simeq 3$. Not surprisingly,  as in the original result obtained by direct integration, the large $|a_s k_F|$ limit is not correctly accounted for.
One difference we have observed however is that the dependence of  $ {U_{pp}(k)}$ given by Eq. (\ref{eq:simplePPSGPS}) close to unitarity remains very smooth and
does not presents the bumps seen in Fig. \ref{fig:GEIAEIunit}.

\subsection{PPS with the  Hugenholtz-van-Hove theorem constraint: illustration with the GPS functional for $k<k_F$.}

The simple PPS approximation has some advantages. Among them, we note that the direct strategy used in previous section
automatically leads the correct low density limit while the resulting single-particle potential $U_{pp}$ has a rather compact expression.
We also saw that it gives quite reasonable behavior  much beyond the perturbative regime.
 One drawback is that the corresponding energy obtained by direct integration through Eq. (\ref{eq:emu}) turns out to be
more complex than typically the GPS and/or APS functionals given by Eqs. (\ref{eq:GPS}) or (\ref{eq:APS}).  For this reason, we explored
a different strategy that consists directly in imposing the constraint (iii),  i.e. the energy obtained when
applying Eq. (\ref{eq:general-HvHthm}) should match a pre-selected expression of the energy (the APS, GPS, GUL, AUL expressions
for instance). In practice, this strategy is much less straightforward
since the chemical potential is imposed whatever the value of
$k_F$ and $a_s$.  It however has the direct advantage that all nice properties that were obtained at the energy level are automatically
incorporated in the single-particle potential.

Our starting point is to pre-suppose that we already know the expression of the energy in terms of $(a_s k_F)$.  As an illustration, we consider
below that the energy should match the GPS energy given by Eq. (\ref{eq:GPS}) obtained by the $pp$ ladder approximation.  From the imposed energy, and using Eq. (\ref{eq:general-HvHthm}), we obtained that the potential at $k=k_F$ should verify:
\begin{eqnarray}
\frac{\mu}{\mu_{FG}} &=&  1+ \frac{U_{pp}(k_F)}{\mu_{FG}}  = 1+ \frac{\dfrac{4}{3\pi}(a_sk_F)}
                     {1- \dfrac{9\pi}{14}(a_sk_F)\Phi_2(1)}
  + \frac{\dfrac{1}{7} (a_sk_F)^2\Phi_2(1)}
                       {\left[1- \dfrac{9\pi}{14}(a_sk_F)\Phi_2(1)\right]^2}
  \nonumber \label{eq:PS-HvHthm}
\end{eqnarray}
where we have used the expression of $\Phi_2(1)$ given by equation (\ref{eq:re-galitskii-LO}) directly recognized
in the chemical potential. From this, we see that imposing the HvH theorem at $k=k_F$ (or $p=1$) gives us a strong guidance on the single-particle potential
for $k\neq k_F$ (or $p\neq1$).   The simplest approximation that could be directly inferred from $\mu$ to obtain the potential consists in
replacing $\Phi_2(1)$ by $\Phi_2(p)$, i.e.:
\begin{eqnarray}\label{eq:Upp-xcst}
  \frac{U_{pp}(k)}{\mu_{FG}}
  & = &
   \frac{\dfrac{4}{3\pi}(a_sk_F)}
                     {1- \dfrac{9\pi}{14}\Phi_2(p)(a_sk_F)}
  + \frac{\dfrac{1}{7}\Phi_2(p) (a_sk_F)^2}
                       {\left[1- \dfrac{9\pi}{14}\Phi_2(p)(a_sk_F)\right]^2}.
                       \label{eq:UGPS}
\end{eqnarray}
The present single-particle potential presents several specific properties:
\begin{itemize}
  \item First, similarly to the simple PPS approach presented in previous section, its expansion to second order in $(a_S k_F)$
  matches the exact result for low density Fermi gas.
  \item It also has automatically a non-divergent behavior in the limit $|a_s k_F| \rightarrow + \infty$.  Due to the HvH theorem constraint, the associated limit is compatible with the value of the associated Berstch parameter, $\xi_{\rm GPS}$ in the present illustration.
  \item The form of the single-particle potential turns out to be slightly more complicated than in the simple PPS approximation,
  Eq. (\ref{eq:simplePPSGPS}). It is worth mentioning however that this form has strong similarities with the single-particle potential
  obtained by $pp$ ladder approximation (see Eq. (\ref{eq:upp})), in particular with the presence of two terms with similar $(a_s k_F)$ dependence
  as in Eq. (\ref{eq:Upp-xcst}).
  \item  Last, we note that the HvH constraint solely does not uniquely define the form of the potential $U_{pp}$. Indeed, we can fulfill this constraint and keep all above mentioned properties using for instance the generalized formed:
 \begin{eqnarray}
  \frac{U_{pp}(k)}{\mu_{FG}}
   &=&    \frac{(4/3\pi)(a_sk_F)}
                     {1- (3\pi/4)\Phi_2(p)X(p)(a_sk_F)} \nonumber \\
&+& \frac{\Phi_2(p)\left[1-X(p)\right] (a_sk_F)^2}
                       {\left[1- (3\pi/4)\Phi_2(p)X(p)(a_sk_F)\right]^2} , \label{eq:UGPSX}
\end{eqnarray}
where the only constraint on $X(p)$ is that $X(1)=6/7$. Again this flexibility should be seen as a positive point since it might be used
to impose additional constraints latter on. In the following, we will generically denotes the potential given by Eq. (\ref{eq:UGPSX}) simply by $U_{{GPS}_X}$ and refer to it as the GPS$_X$ approximation, while the case $X(p)=6/7$, leading to Eq. (\ref{eq:Upp-xcst}) is simply called GPS approximation for the single-particle potential and is noted as $U_{GPS}$. Unless specified, results presented below will be obtained in the GPS
approximation.
  \end{itemize}

Equivalent strategy can be followed starting from the APS approximation.  Imposing the HvH theorem consistent with the APS approximation for
the energy, we end-up with the following expression for the single-particle potential:
\begin{eqnarray}
  \frac{U_{APS_X}(k)}{\mu_{FG}} & = &
  \frac{(2/9\pi) a_sk_F }{\left[1- (9\pi/4)\Phi_2(p)X(p)a_sk_F\right]^2  +\left[ (5/24) a_sk_F\right]^2}
  \nonumber \\ \label{eq:def-APSX}
   &+&
  \frac{16}{3 \pi } \arctan \left(\frac{(5/24)a_sk_F}{1- (9\pi/10)\Phi_2(p)\left[1-X(p) \right]a_sk_F}\right).
\end{eqnarray}
This approximation is called APS$_{X}$ hereafter. The only constraint on the function $X(p)$ is now $X(p=1)=2/7$. If we assume
that this function is constant for all $p$, approximation called simply APS hereafter, we end up with the  single-particle potential:
\begin{eqnarray}
  \frac{U_{APS}(k)}{\mu_{FG}}  &=&
  \frac{(2/9\pi) a_sk_F }{\left[1- (9\pi/14)\Phi_2(p)a_sk_F\right]^2  +\left[ 5 a_sk_F/24\right]^2}
  \nonumber \\
  &+&
  \frac{16}{3 \pi } \arctan \left(\frac{5a_sk_F/24}{1- (9\pi/14)\Phi_2(p)a_sk_F}\right).
  \label{eq:UAPS}
\end{eqnarray}
Again, all constraints (i-iii) are respected and whatever the explicit form of $X(p)$, in the infinite scattering
length limit, we obtain the APS value for the Bertsch parameter.

\begin{figure}[htbp]
\begin{center}
 \includegraphics{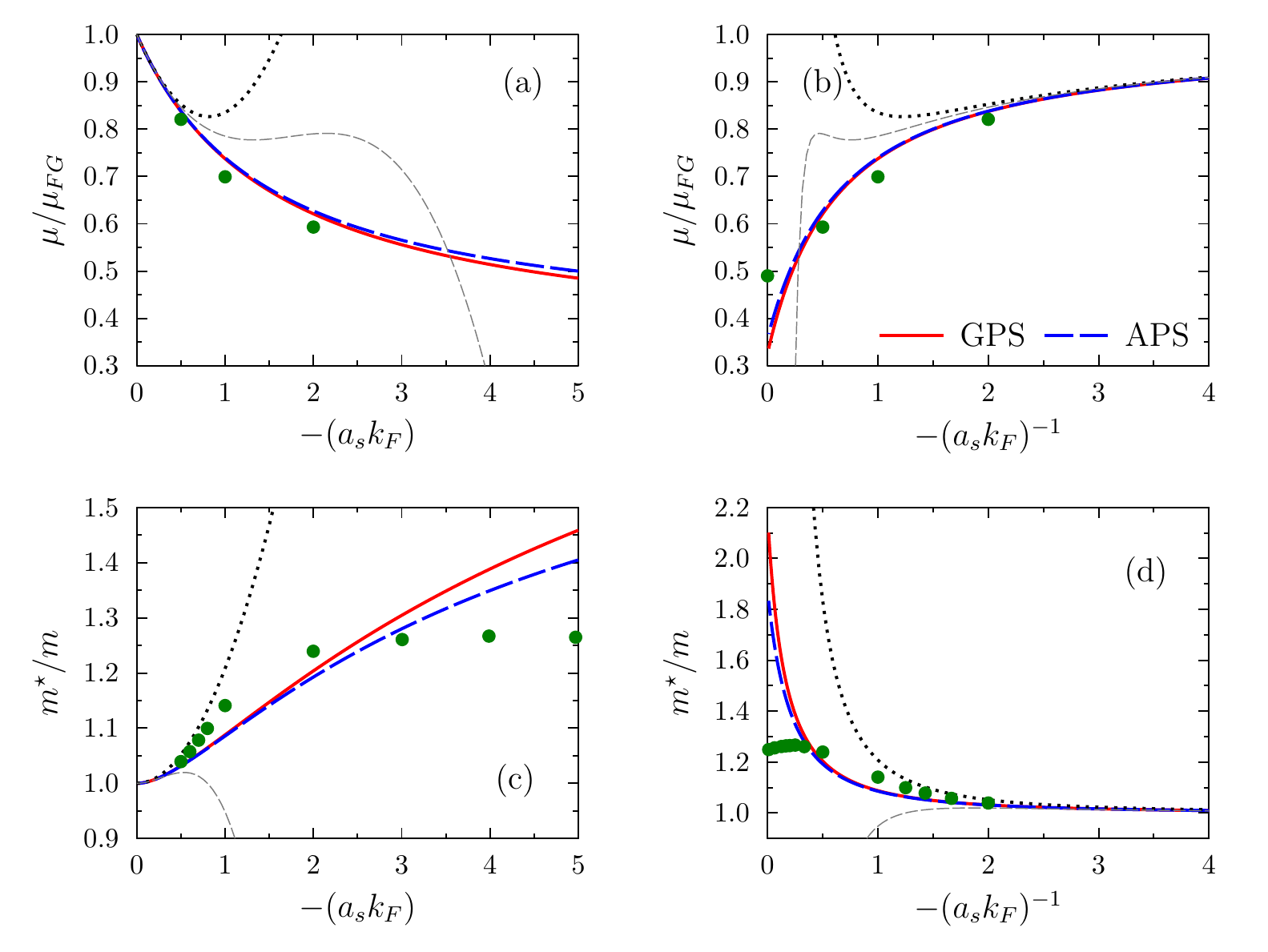}
\caption{Chemical potential and effective mass as a function of $(a_s k_F)$ [panels (a) and (c)] or $-(a_s k_F)^{-1}$
[panels (b) and (d)] obtained with the GPS (red solid line) and APS
(blue dashed line) approximations. In panels (a) and (c), result of the second order (Galitskii formula) and third order expansion \cite{Pla03}
in $(a_s k_F)$ are shown with back dotted and thin gray dashed lines respectively. The green circles correspond to the result of the BHF calculations of Ref.  \cite{Dog14}.}
\label{fig:mumstarPPSas}
\end{center}
\end{figure}

The chemical potential and effective mass dependence in $(a_s k_F)$ obtained with the consistent GPS and APS scheme are displayed in Fig. \ref{fig:mumstarPPSas} respectively as a function of $-(a_s k_F)$ or $-(a_s k_F)^{-1}$.  The conclusion are similar as for the simple GPS case presented previously, i.e. the low density limit (respectively Lee-Yang and Galitskii formula) are properly reproduced by construction. The BHF results is reproduced qualitatively up to $-(a_s k_F) \simeq 3$ while the perturbative expansion breaks down
around $-(a_s k_F) = 0.5$. We note however that the result of the consistent APS approximation is slightly worse compared to the original AEI as far as the chemical potential is concerned.

\section{Quadratic and quartic approximation for the single-particle potential}

Our targeted goal here is to provide DFT inspired by the many-body resummation technique presented above. The clear advantage
to start from the self-energy level instead of the energy itself is that direct connections can be made between the self-energy and the
Fermi Liquid theory. This was illustrated previously with the chemical potential and the effective mass. Such quantities are also standardly
obtained with Energy Density Functionals for instance used in the nuclear physics context, like Skyrme or Gogny EDFs. Empirical functionals, especially the functionals derived using Skyrme like contact interactions lead to very simple single-particle potentials (see discussion below) with polynomial dependence in $k$. For instance, the original parameterization proposed in \cite{Vau72} leads simply
to quadratic dependence of the single-particle potential in infinite matter.  Novel generations of Skyrme EDF have been proposed leading to
quartic or higher-order dependence in the momentum \cite{Rai11,Dav14,Dav18}.
The justification that such simple approximation can contain
important physical aspects can be found in \cite{Fet71a}. The different single-particle potentials obtained in previous section presents rather complex
density and momentum dependence. However, starting from the PPS approximation, one might obtain a systematic polynomial expansion
to a given order in $k$.
For this, we approximate the self-energy obtained by assuming a polynomial form. We introduce the following expansion:
\begin{eqnarray}
U(k) \simeq U_0(k_F) +  U_2(k_F)(k/k_F)^2 +  U_4(k_F)(k/k_F)^4 + \cdots \label{eq:expU}
\end{eqnarray}
This polynomial expansion, truncated at an appropriate order will not only be useful to make contact with empirical
density functional theory but will also enable to obtain practical DFT for finite systems based on the present approach.
\begin{figure}\centering
 \includegraphics{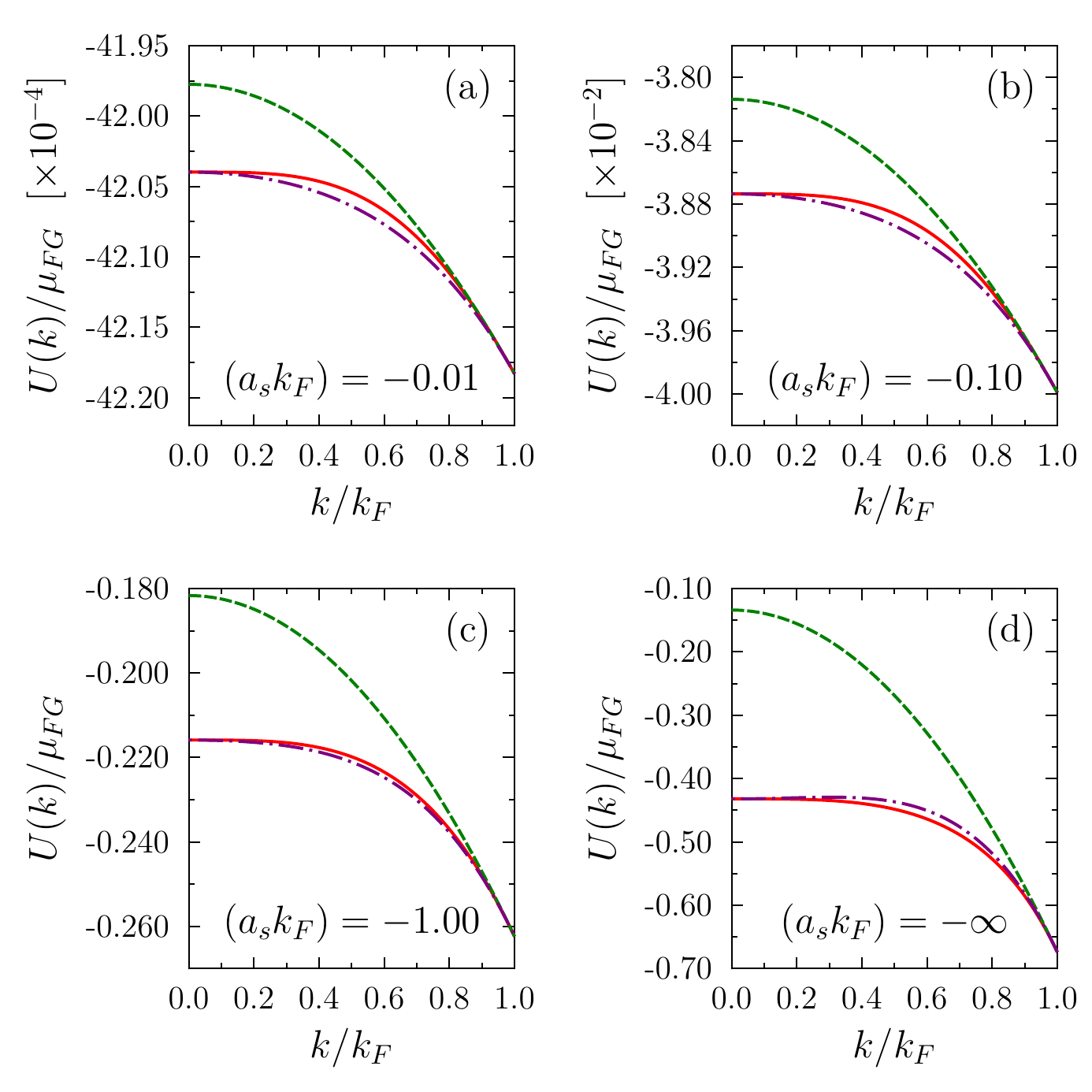}
  \caption{Momentum dependence of the approximated self-energy for $-(a_sk_F) = 0.01$ (a), 0.1 (b), $1$ (c) and $\infty$ (d) obtained with the GPS approximation (red solid line).
  In each panel, the corresponding quadratic or quartic approximation are shown respectively by the green dashed line
 and  purple long-short dashed line.}
  \label{fig:UquasGPS}
\end{figure}
\begin{figure}\centering
 \includegraphics{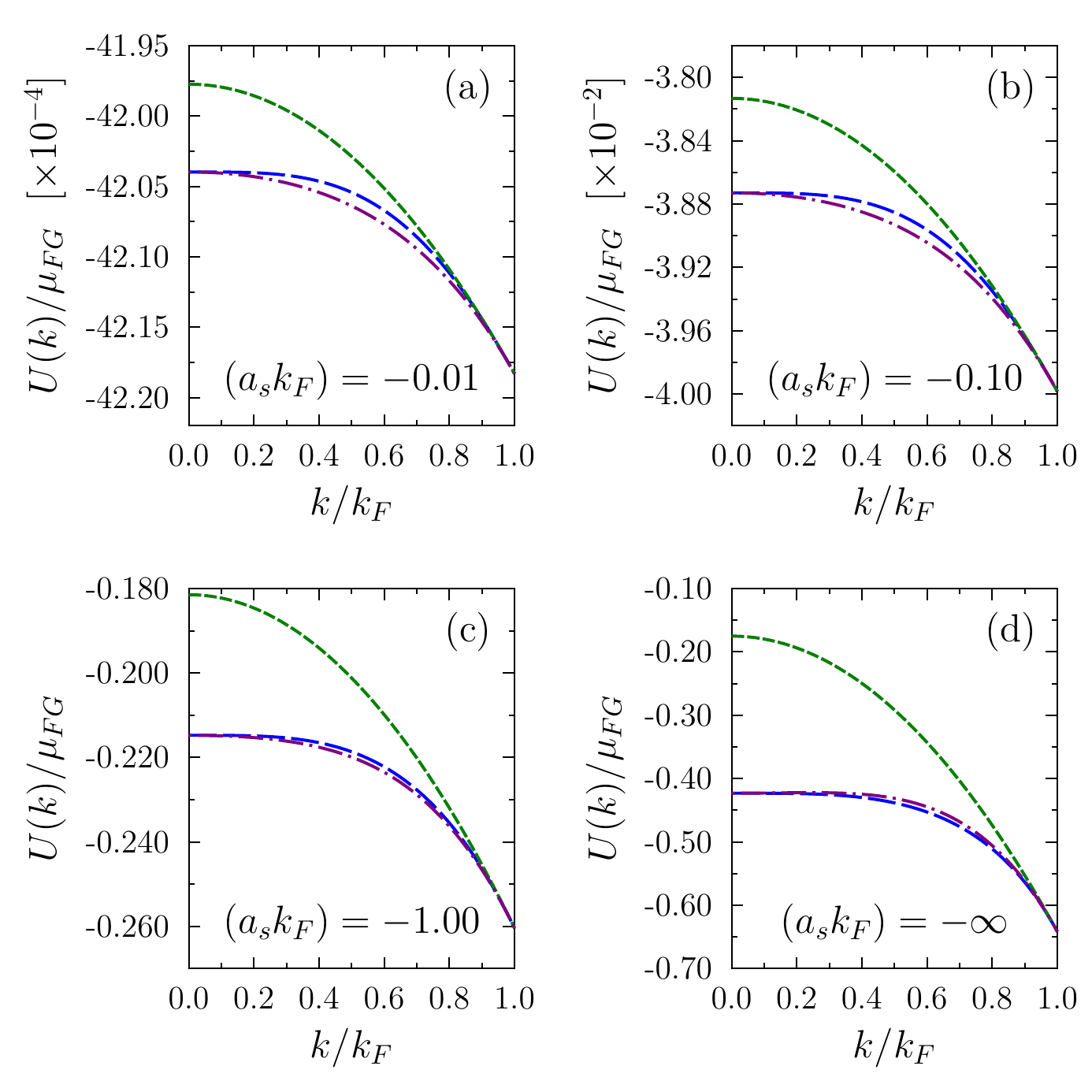}
  \caption{Same as Fig. \ref{fig:UquasGPS} for the APS approximation. The reference APS curve is now shown in blue long-dashed line.}
  \label{fig:UquasAPS}
\end{figure}

\subsection{Quadratic approximation for the self-energy}

The simplest  approximation that is also certainly the most highlighting one for the present discussion  is to
consider quadratic single-particle potential, i.e. keeping only $U_0$ and $U_2$ in Eq. (\ref{eq:expU}). Only two
constraints are then needed to obtain $U_0$ and $U_2$. One possibility is to impose that some of the quasi-particle
properties are exactly recovered. For instance, imposing the chemical potential and the effective mass to be the same as the original
ones obtained with one of the PPS approximation leads to:
\begin{eqnarray}
1 + \frac{U_0(k_F)}{\mu_{FG}} + \frac{U_2(k_F)}{\mu_{FG}} &=&\frac{\mu(k_F)}{\mu_{FG}}  , ~~~~1 + \frac{U_2(k_F)}{\mu_{FG}} =  \frac{m}{m^*(k_F)} , \nonumber
\end{eqnarray}
giving finally:
\begin{equation}
  \frac{U(k)}{\mu_{FG}}
  =\left[\frac{\mu(k_F)}{\mu_{FG}} - \frac{m}{m^*(k_F)}\right] + \left[\frac{m}{m^*(k_F)}-1\right] \left(\frac{k}{k_F}\right)^2. \label{eq:Uquad}
\end{equation}
Since they are used as constraint, the present method automatically insures that the quasi-particle properties are preserved
even if a simplified expansion is used for $U(k)$. Since the chemical potential is also constrained,  due to the relationship (\ref{eq:emu}),
the energy of the system will also be preserved. Said differently, the GPS (resp. APS) approximation for the self-energy combined
with the polynomial approximation of $U(k)$ will lead to the GPS (resp. APS)  reference energy given by Eq. (\ref{eq:GPS})  (resp. (\ref{eq:APS})).
This does not necessarily implies that the potential are similar, however by construction, they should become  identical as $k$
becomes close to $k_F$. We compare in Fig. \ref{fig:UquasGPS} and \ref{fig:UquasAPS} for the GPS and APS cases,
the original GPS and APS potentials given by Eqs. (\ref{eq:UGPS}) and (\ref{eq:UAPS}) respectively,
with their quadratic approximations for different values of $(a_s k_F)$. We clearly see in this figure that the value and the slope
of $U(k)$ at $k=k_F$ that are respectively linked to the chemical potential and the effective mass are identical. However some differences are observed as $k/k_F$ goes to zero. This also implies that some deviations occur with  the second-order expansion of
the self-energy given by Eq. (\ref{eq:Usecond}) when the quadratic form of the potential is used in the low density limit. However,
this approximation still leads to the proper behavior given by Eq. (\ref{eq:mmstarNLO}) in this limit.

\begin{figure}\centering
     \includegraphics{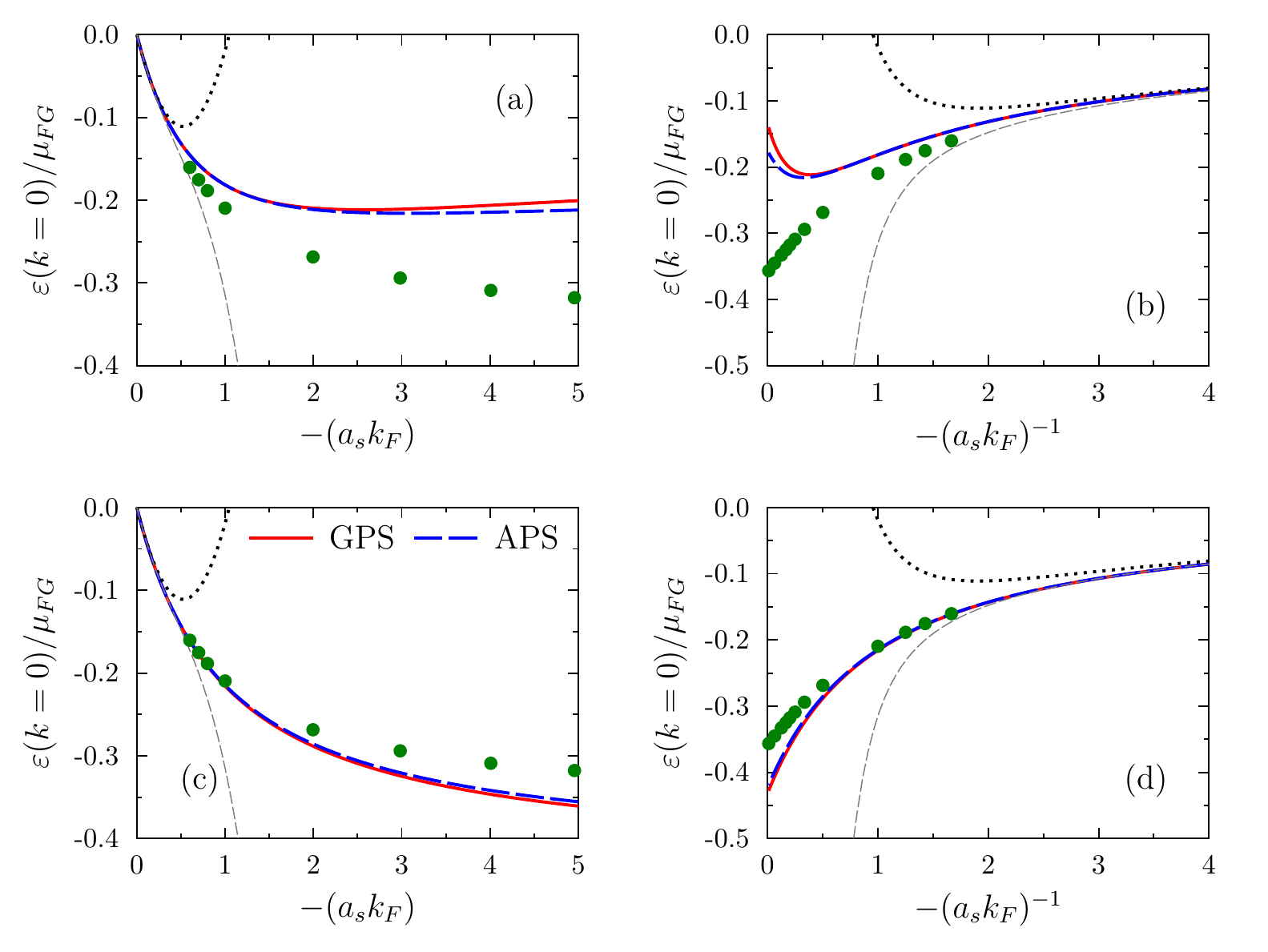}
     \caption{The single-particle energy $\varepsilon(k=0)$ at zero-momentum obtained from the GPS (red solid line) and APS (blue dashed line)
     single-particle potentials as function of $-(a_sk_F)$ [panels (a) and (c)] or  $-(a_sk_F)^{-1}$ [panels (b) and (d)]. In panels (a) and (b) (resp. (c) and (d))
     the results obtained using a quadratic (resp. quartic) approximation are shown in all panels by red solid line (GPS case)
     and blue dashed line (APS case). Note that in the quartic approximation, by construction (see text) the value
     of  $\varepsilon(k=0)$ is equal to the original APS or GPS value. The black dotted line and gray dashed line correspond respectively
     to the value obtained from a quartic or quadratic approximation starting from the second order (Eq. (\ref{eq:Usecond})) or third order \cite{Pla03}
     expansion of the self-energy in $(a_s k_F)$ respectively. In all panels, the green circles correspond to the BHF results of Ref. \cite{Dog14}} \label{fig:esp0}
\end{figure}

The difference observed between the original GPS or APS approximations and the quadratic expansion given by Eq.
(\ref{eq:Uquad}) is further illustrated in Fig. \ref{fig:esp0} where we display the single-particle energy obtained in different cases
at $k=0$.

\subsection{Quartic approximation for the self-energy}

The agreement between the original potential and the polynomial expansion can be significantly improved simply by going to a quartic
form of the potential, i.e. by truncating the expansion (\ref{eq:expU}) to the next order.  Then, the three parameters can be adjusted
by adding to the chemical potential and effective mass constraint, the additional  constraint that the value of $U(k)$ at $k=0$
is identical to the one of the original GPS and APS approximation. We see in Fig. \ref{fig:UquasGPS} and  \ref{fig:UquasAPS} respectively for the GPS and
APS cases,  that the quartic approximation significantly improves
the global shape of the potential compared to the original PPS approximation.

\section{Summary}

In the present work, we explore the possibility to start from a well-defined many-body approach based on the diagrammatic resummation technique
and obtain approximate expressions  both for the energy and/or the self-energy. While the present work is mainly focused on this topic, the ultimate
goal is to use these approximate expression  as a guidance for proposing new DFT for Fermi systems beyond the perturbative regime.
At the self-energy level, two simplifications are considered that might help in the DFT context. We first propose to use a partial phase-space
approximation leading to simpler density dependence of the self-energy  and ultimately of the associated energy. We show that the partial phase-space
average  can be made either by imposing the form of the energy simultaneously or not. In the former case, the constraint on the energy
is made using the  Hugenholtz-van-Hove theorem consistency. If the energy is not used as a constrained, the associated self-energy expression turns out to
be simpler. In all cases, the self-energy of low density Fermi gas is properly recovered while a non-diverging limit is reached when
$|a_s k_F| \rightarrow + \infty$ contrary to truncated perturbative many-body framework.
It is found in general that the combination of a resummation technique with phase-space average approximation can be used in a wider range of densities compared
to many-body perturbation theory. We note however that without any adjustment, the functionals are not predictive in the unitary regime.

Besides the simplification introduced by the phase-space averages, we note
in general that the resulting energy and/or self-energy present both  smoother behaviors that turn out to be qualitatively closer to the behavior calculated through Brueckner Hartree-Fock  approach for non superfluid systems compared to a direct use of the resummation approach without PS approximation.

Guided by some phenomenological arguments commonly used for nuclei and also by the success of simple empirical functionals  like the Skyrme DFT in nuclear systems,
the single-particle potential is further simplified by assuming that the single-particle potential can be approximated by a quadratic or quartic polynomial in $k$. This second approximation
is made in such a way that quasi-particle properties are preserved for all $(a_s k_F)$. Again, this has the advantage that the low density limit of the chemical potential
or the effective mass identifies automatically respectively with the Lee-Yang and Galitskii expression. We finally briefly discuss how the possibility
to have analytical density dependent expressions of these quasi-particle properties can serve to design new DFTs. More generally, in view
of the recent scientific emulation that followed the use of resummed formula for the energy \cite{Lac16a,Yan16,Gra17,Gra18}, the approximate expressions
obtained in the present work can also be a strong guidance to obtain semi-empirical or non-empirical functionals constrained at low density or at unitary or both.

As an illustration of this guidance, let us assume that we start from one of the parametrized forms of the chemical potential and effective mass obtained
in the present work. We can then first rewrite the energy of the system in infinite system as:
\begin{eqnarray}
\frac{E}{N} &=& \frac{3}{5} \frac{\hbar^2 k^2_F}{2m^*(\rho)} + W(\rho) . \nonumber
\end{eqnarray}
Following Ref. \cite{Lip08}, we can show that this energy can be obtained from a density functionals valid in both finite and infinite systems
of the form:
 \begin{eqnarray}
E(\rho, \tau) &=& \int \left\{ \frac{\tau({\bf r})}{m^*(\rho)} + \rho({\bf r}) W[\rho({\bf r})]  \right\} d {\bf r}
\end{eqnarray}
where the local density $\rho$ and the  kinetic density $\tau$  entering in this equation are linked with the Kohn-Sham states $\{\varphi_i \}$
through $\rho({\bf r}) = \sum_{i=1}^N |\varphi_i ({\bf r})|^2$ and $\tau({\bf r}) =   \sum_{i=1}^N |\nabla\varphi_i ({\bf r})|^2$. One might notice
in particular that in the present approach, corrections beyond the HF are automatically incorporated in both the potential and effective mass terms.
This direction will be further explored in the near future.

The use of diagrammatic resummation leads to rather complex expressions in general.
For this reason, we focused here the discussion on a rather simple case of a non-superfluid system with only one low energy
constant, the s-wave scattering length focusing on the on-shell self-energy. In addition, an extension to include off-shell effects would be
a priori desirable especially to describe the E-mass \cite{Jeu76a,Mah93a}.
Another natural extension of the present work is to include also the s-wave effective range $r_e$ and/or
p-wave scattering volume. The functional proposed in Ref. \cite{Lac16a} and further discussed in \cite{Lac17a,Bou18} already
incorporate the effect of the effective range on the energy density functional.
At leading order, one could simply add the present functional, however a proper treatment of the possible interference effects as well
as effects beyond the leading order is needed if for instance $(r_e k_F)$ becomes large. This case happens for instance in nuclear system
at saturation density. We did not consider here the treatment of the effective range together with large scattering length, however we mention
that the work of Refs. \cite{Lac11,Kai12} can be use as a starting point.

Another important extension would be to include the effect of pairing correlations. As we mentioned in the introduction, a prerequisite  for the
the present study is the possibility to obtain a compact expression of the energy after summing up selected diagrams to all orders in perturbations.
By itself developing a perturbative approach on top of a quasi-particle vacua is possible (see for instance \cite{Lac12,Rip17}). However, even at second
order, by replacing the 2p-2h energy by the 4 quasi-particle, the complexity of the integrals to be made increases significantly. As far as we know,
such problem has not be solved analytically. An alternative to the analytical approach using a DFT guided by the present work would be simply
to add a posteriori a pairing energy to the DFT.  This procedure is standardly used for nuclei and would at least allow to extend to the so-called
Superfluid LDA (SLDA) approach of Ref. \cite{Bul07} away from unitarity.

\ack{We thank H. F. Arellano, J. Bonnard, M. Grasso, A. Gezerlis,  C.-J. Yang for useful discussions
at different stage of this work.  This project has received funding from the European Union's Horizon 2020 research and innovation
programme under Grant Agreement No. 654002.
}

\appendix

\section{Useful definitions and integrals}
\label{app:eresum}

In the main text, several quantities are written as integrals over the phase-space.
For the sake of completeness, the different functions defined in the text
as well as the different variable are summarized in this appendix. Note that a complete derivation of
all equations can be found in Ref. \cite{Bou19}

Our starting point is the interaction matrix elements (\ref{eq:C0}) written in momentum space
as $\langle \bm{k_1}\bm{k_2} | V_{\rm EFT} |  \bm{k_3}\bm{k_4} \rangle $ (note that, the spin is implicit and will lead
to factors in the energy).  Different functions appearing in the integrals for the energy and/or self-energy
are written as a function of $s = |\bm{s}|$ and $t=|\bm{t}|$ where $\bm{s}$   and $\bm{t}$ are vectors defined through:
\begin{eqnarray}
\bm{s} = \frac{\bm{k_1}+\bm{k_2}}{2k_F},
~~
\bm{t} = \frac{\bm{k_1}-\bm{k_2}}{2k_F}. \label{eq:st}
\end{eqnarray}

After proper treatment of the UV divergence and after averaging over vectors relative angles, it could be shown that the energy
take the form (\ref{eq:GEI}) and (\ref{eq:AEI}),  where $I$, $F$ and $R$ are given respectively by \cite{Kai11,Kai13}:
\begin{eqnarray*}
  I(s,t) &=&  t  \displaystyle\min\left[1;  \left|\frac{1-s^2-t^2}{2st}\right| \right]  \\
  Y(s,t) &=&  1+s+t\ln\left|\frac{1+s-t}{1+s+t}\right| + \frac{1-s^2-t^2}{2s}\ln\left|\frac{(1+s)^2-t^2}{1-s^2-t^2}\right| \\
	F(s,t) &=& Y(s,t) + \Theta(s-1)Y(-s,t) \\
	R(s,t) &=& Y(s,t) + Y(-s,t)
\end{eqnarray*}

\subsection{Integrals used for phase-space average of the energy}
\label{eq:integrals}

The following integral are used to obtain the Phase-space averaged resummed  expression for the energy:
\begin{eqnarray}
\langle \langle 1 \rangle \rangle &=&   \int_0^1 s^2 ds \int_0^{\sqrt{1-s^2}} tdt  =  \frac{1}{15}
  \nonumber \\
\langle \langle I \rangle \rangle&=&   \int_0^1 s^2 ds \int_0^{\sqrt{1-s^2}} tdt I(s,t)  =  \frac{1}{72}
  \nonumber\\
\langle \langle I R\rangle \rangle &=&  \int_0^1 s^2 ds \int_0^{\sqrt{1-s^2}} tdt I(s,t)R(s,t)  = \frac{1}{72}\times\frac{6}{35}(11-2\ln2)
\nonumber\\
\langle \langle I F\rangle \rangle &=&  \int_0^1 s^2 ds \int_0^{\sqrt{1-s^2}} tdt I(s,t)F(s,t)  = \frac{1}{72}\times\frac{6}{35}(11-2\ln2)
\nonumber
\end{eqnarray}

\section{Functions used in the self-energy}
\label{app:self}

The two functions entering in the second-order self-energy, Eq. (\ref{eq:self12}) are given by \cite{Gal58}:
\begin{eqnarray}
	\Phi_2(p) &=& \frac{4}{15\pi^2}\frac{1}{p}
	\Bigg\{11p + 2p^5\ln\left|\frac{p^2}{p^2-1}\right|  -10(p^2-1)\ln\left|\frac{p+1}{p-1}\right| \Big\}
        \nonumber \\
       &-& \frac{8}{15\pi^2}\frac{|2-p^2|^{5/2}}{p} \Bigg\{\Theta(2-p^2)\ln\left|\frac{1+p\sqrt{2-p^2}}{1-p\sqrt{2-p^2}}\right| \nonumber\\
       && \qquad\qquad \qquad \!+  \Theta(p^2-2)\cot^{-1}\sqrt{p^2-2} \Bigg\},  \label{eq:re-galitskii}
	\\
	\Omega_2(p) & = & \Theta(1-p)\frac{(1-p^2)^2}{2\pi} \nonumber \\ &-&\Theta(p-1)\frac{2}{15\pi^2}\frac{1}{p}\left\{5p^2-7+2(2-p^2)^{5/2}\Theta(\sqrt{2}-p) \right\}.
  \label{eq:im-galitskii}
\end{eqnarray}
Here, we use the Heaviside step function $\Theta(x)$ to shorten the notations.

\subsection{Functions entering in the expression of the resummed self-energies}
\label{app:selfresum}

{\bf Particle-particle and hole-hole ladder diagrams resummation:}
In this case, we write the self-energy $\Sigma(k) = U(k)+{\rm i} W(k) $ as:
\begin{eqnarray}
  \frac{\Sigma (k)}{\mu_{FG}} & =&
  \Theta(k_F-k)\int_0^{1}s^2 ds \int_0^{\sqrt{1-s^2}} \hspace{-15pt}tdt ~ \bigg[ \mathcal{U}(s,t,p)+{\rm i} \mathcal{W}(s,t,p) \bigg]
  \nonumber \\ &+&
  \Theta(k-k_F)\int_0^{(1+p)/2}s^2 ds \int_0^{(1+p)/2} tdt ~\bigg[ \mathcal{U}^\prime(s,t,p)+{\rm i} \mathcal{W}^\prime(s,t,p) \bigg] \nonumber
\end{eqnarray}
The different functions are given by ($k<k_F$):
\begin{eqnarray} \label{eq:UKaiser-below}
 \frac{  \mathcal{U}(s,t,p)  }{\mu_{FG}} &=&
 \frac{16(a_sk_F)^2 I_*(s,t) \widehat{R}(s,t,p) + (a_sk_F)\widehat{I_*}(s,t,p)\left[\pi-(a_sk_F)R(s,t)\right]}{\left[\pi-(a_sk_F)R(s,t)\right]^2 + \left[(a_sk_F)\pi I(s,t)\right]^2} \nonumber \\
 &-& 16\widehat{R}(s,t,p) \delta\left(\pi/(a_sk_F) - R(s,t)\right) \Theta(1-s^2-t^2), \\
  \frac{  \mathcal{W}(s,t,p)  }{\mu_{FG}}   &=&
   \frac{16\pi (a_sk_F)^2 \left[\widehat{I_*}(s,t,p) - \widehat{I}(s,t,p) \right] I_*(s,t) }{\left[\pi-(a_sk_F)R(s,t)\right]^2 + \left[(a_sk_F)\pi I(s,t)\right]^2}.
\end{eqnarray}
and ($k>k_F$):
\begin{eqnarray}
 \frac{  \mathcal{U}^\prime(s,t,p)  }{\mu_{FG}} &=&
 \frac{16(a_sk_F)^2 I_*(s,t) \widehat{R}(s,t,p) + (a_sk_F)\widehat{I_*}(s,t,p)\left[\pi-(a_sk_F)R(s,t)\right]}{\left[\pi-(a_sk_F)R(s,t)\right]^2 + \left[(a_sk_F)\pi I(s,t)\right]^2}
   \nonumber \\
 &-&16\widehat{R}(s,t,p) \delta\left(\pi/(a_sk_F) - R(s,t)\right) \Theta(1-s^2-t^2), \\
 \frac{  \mathcal{W}^{\prime}(s,t,p)  }{\mu_{FG}} &=&
  -\frac{16\pi(a_sk_F)^2 \widehat{I_*}(s,t,p)I(s,t)\Theta(s^2+t^2-1) }{\left[\pi-(a_sk_F)R(s,t)\right]^2 + \left[(a_sk_F)\pi I(s,t)\right]^2}.
\end{eqnarray}
Note that, these results are equivalent those given by Eqs. (18--21) of Ref. \cite{Kai13} except that the sign convention for the scattering length
is different (i.e. $a_s \leftrightarrows - a_s$).

\noindent{\bf Particle-particle ladder diagrams only  resummation:}
The resummed self-energy $\Sigma_{pp}(k) = U_{pp}(k)+{\rm i} W_{pp}(k) $ is now given by:
\begin{eqnarray}
  \frac{\Sigma_{pp} (k)}{\mu_{FG}} & =&
  \Theta(k_F-k)\int_0^{1}s^2 ds \int_0^{\sqrt{1-s^2}} \hspace{-15pt}tdt ~ \bigg[ \mathcal{U}_{pp}(s,t,p)+{\rm i} \mathcal{W}_{pp}(s,t,p) \bigg]
  \nonumber \\ &+&
  \Theta(k-k_F)\int_0^{(1+p)/2}s^2 ds \int_0^{(1+p)/2} tdt ~\bigg[ \mathcal{U}^\prime_{pp}(s,t,p)+{\rm i} \mathcal{W}^\prime_{pp}(s,t,p) \bigg] .\nonumber
\end{eqnarray}
The different functions are given by ($k<k_F$):
\begin{eqnarray} \label{eq:defUpp-lower}
 \frac{  \mathcal{U}_{pp}(s,t,p)  }{\mu_{FG}} &=&
   \frac{16(a_sk_F)^2\widehat{F}(s,t,p)I_*(s,t)}{\left[\pi - (a_sk_F)F(s,t) \right]^2} + \frac{16(a_sk_F)\widehat{I_*}(s,t,p)}{\pi - (a_sk_F)F(s,t) }
\label{eq:upp}  \\
  \frac{  \mathcal{W}_{pp}(s,t,p)  }{\mu_{FG}} & = &
   \frac{16\pi(a_sk_F)^2\left[\widehat{I_*}(s,t,p)-\widehat{I}(s,t,p) \right]I_*(s,t)}{\left[\pi - (a_sk_F)F(s,t) \right]^2}
\end{eqnarray}
and ($k> k_F$):
\begin{eqnarray}\label{eq:defUpp-upper}
\frac{  \mathcal{U}'_{pp}(s,t,p)  }{\mu_{FG}} &=&
   \frac{16(a_sk_F)^2\widehat{F}(s,t,p)I_*(s,t)}{\left[\pi - (a_sk_F)F(s,t) \right]^2}
 + \frac{16(a_sk_F)\widehat{I_*}(s,t,p)\Theta(1-s^2-t^2)}{\pi - (a_sk_F)F(s,t) }  \nonumber \\
 &+&  \frac{16(a_sk_F)\left[\pi - (a_sk_F)F(s,t) \right]\widehat{I_*}(s,t,p)\Theta(s^2+t^2-1)}{\left[\pi - (a_sk_F)F(s,t) \right]^2 + \left[(a_sk_F)\pi I(s,t) \right]^2}
  \nonumber \\
\frac{  \mathcal{W}'_{pp}(s,t,p)  }{\mu_{FG}} &=&   -
  \frac{16\pi(a_sk_F)^2\widehat{I_*}(s,t,p)I(s,t)\Theta(s^2+t^2-1)}{\left[\pi - (a_sk_F)F(s,t) \right]^2 + \left[(a_sk_F)\pi I(s,t) \right]^2}
\end{eqnarray}
The above quantities uses many more functions that are listed below:
\begin{eqnarray}
I_*(s,t) &=&  I(s,t) \Theta(1-s^2-t^2) \nonumber \\
\widehat{I}(s,t,p) &=& \frac{1}{sp}\Theta(s+t-p)\Theta(p-|s-t|)~\mathrm{sign}\left((1+p^2)/2-s^2-t^2\right) \nonumber \\
\widehat{I_*}(s,t,p) &=& \frac{1}{sp}\Theta(s+t-p)\Theta(p-|s-t|)\Theta\left((1+p^2)/2-s^2-t^2\right) \nonumber \\
\widehat{R}(s,t,p) &=&  \frac{1}{sp} \ln \left|\frac{(s+p)^2-t^2}{(s-p)^2-t^2} \right|  \nonumber \\
  \widehat{F}(s,t,p) & = & \frac{1}{sp}\left\{ \Theta(2s-1)\Theta(2s-1-p)\ln\left|\frac{(s+p)^2-t^2}{(s-p)^2-t^2} \right| \right.
  \nonumber \\
  &+&\left.\Theta(p-|2s-1|)\ln\left|\frac{(s+p)^2-t^2}{(1+p^2)/2-s^2-t^2}\right|\right\}\Theta(1-p) \nonumber \\
  &+& \frac{1}{sp}\left\{ \Theta(p-2s-1)\ln\left|\frac{(s+p)^2-t^2}{(s-p)^2-t^2} \right| \right.
  \nonumber \\
  &+&\left.\Theta(2s+1-p)\ln\left|\frac{(s+p)^2-t^2}{(1+p^2)/2-s^2-t^2}\right|\right\}\Theta(p-1) \nonumber
\end{eqnarray}

\subsection{Partial Phase-Space average for different functions appearing in the self-energy}
\label{app:pps}

Starting from the expression (\ref{eq:upp}), one can expand the potential as in Eq. (\ref{eq:expand-Upp}) where the $  \Phi_n(p)$ are given by:
\begin{eqnarray}
  \Phi_n(p) & = &\frac{16}{\pi^n} \langle \widehat{I}_*F^{n-1} + (n-1)  F^{n-2} \widehat{F}I_* \rangle^<_p  . \label{eq:ppsphin}
\end{eqnarray}
The PPS useful in the article are given by (note that some of them are independent on $p$):
\begin{eqnarray}
  \langle \widehat{I}_* \rangle^<_p  =  \frac{1}{12} , ~~~
  \langle 1 \rangle^<_p              = \frac{1}{15}, ~~~
  \langle I_* \rangle^<_p            = \frac{1}{72}, ~~~
  \left\langle \widehat{I}_*F  + \widehat{F}I_* \right\rangle^<_p   & =& \frac{\pi^2}{16} \Phi_2(p)\nonumber
\end{eqnarray}

\end{document}